\documentclass[longauth]{aa}
\usepackage[utf8]{inputenc}
\bibpunct{(}{)}{;}{a}{}{,}
\usepackage{graphicx,natbib,url,twoopt}
\usepackage[T1]{fontenc}
\usepackage{rotating,booktabs}
\usepackage{float}
\usepackage{adjustbox}
\usepackage{xcolor}

\usepackage{amsmath} 
\title{The CORALIE survey for southern extrasolar planets XVIII 
\thanks{Based on observations collected with the CORALIE spectrograph mounted on the 1.2 m Swiss telescope at La Silla Observatory and with the HARPS spectrograph on the ESO 3.6 m telescope at La Silla (ESO, Chile).}
\thanks{The radial velocity measurements and additional data products discussed in this paper are available on the DACE web platform  at https://dace.unige.ch/radialVelocities. See the appendix for a direct link to the individual target data products.}}
\subtitle{Three new massive planets and two low mass brown dwarfs at separation larger than 5~AU}

\author{E. L. Rickman \inst{1}, D. S\'{e}gransan \inst{1}, M. Marmier\inst{1}, S. Udry\inst{1},  F. Bouchy\inst{1}, C. Lovis\inst{1}, M. Mayor\inst{1}, F. Pepe\inst{1}, D. Queloz\inst{1}, N. C. Santos\inst{2,3}, 
R. Allart \inst{1},
V. Bonvin\inst{4}, 
P. Bratschi \inst{1},
F. Cersullo \inst{1},
B. Chazelas \inst{1},
A. Choplin\inst{1}, 
U. Conod \inst{1},
A. Deline \inst{1},
J.-B. Delisle \inst{1},
L. A. Dos Santos \inst{1},
P. Figueira\inst{5,2},
H. A. C. Giles \inst{1},
M. Girard \inst{1},
B. Lavie \inst{1},
D. Martin \inst{1,6,7},
F. Motalebi \inst{1},
L. D. Nielsen \inst{1},
H. Osborn \inst{8,9},
G. Ottoni\inst{1},
M. Raimbault \inst{1},
J. Rey \inst{1,10},
T. Roger \inst{1,11},
J. V. Seidel \inst{1},
M. Stalport \inst{1},
A. Suárez Mascareño \inst{1,12},
A. Triaud \inst{13},
O. Turner \inst{1},
L. Weber \inst{1},
A. Wyttenbach\inst{1,14}
}
\institute{D\'epartement d'astronomie, Universit\'{e} de Gen\`{e}ve, 51 ch. des Maillettes, 1290 Versoix, Switzerland
\email{emily.rickman@unige.ch} 
 \and
 Instituto de Astrof\'isica e Ci\^encias do Espa\c{c}o, Universidade do Porto, CAUP, Rua das Estrelas, 4150-762 Porto, Portugal
\and
Departamento de F\'isica e Astronomia, Faculdade de Ci\^encias, Universidade do Porto, Rua do Campo Alegre, 4169-007 Porto, Portugal
 \and
Institute of Physics, Laboratory of Astrophysics, École Polytechnique Fédérale de Lausanne (EPFL), Observatoire de Sauverny, 1290, Versoix, Switzerland
  \and
European Southern Observatory, Alonso de Córdova 3107, Vitacura, Santiago, Chile
 \and
 Fellow of the Swiss National Science Foundation
 \and
Department of Astronomy and Astrophysics, University of Chicago, 5640 South Ellis Avenue, Chicago, IL 60637, USA
 \and
Department of Physics, University of Warwick, Coventry CV4 7AL, UK
\and
Aix Marseille Universit\'{e}, CNRS, LAM (Laboratoire d’Astrophysique de Marseille) UMR 7326, F-13388, Marseille, France
\and
Carnegie Institution for Science, Las Campanas Observatory, Casilla 601, Colina El Pino S/N, La Serena, Chile
\and
Physikalisches Institut, Universit{\"a}t Bern, Gesellschaftsstr. 6, 3012 Bern, Switzerland
\and
Instituto de Astrof\'{i}sica de Canarias, E-38205 La Laguna, Tenerife, Spain
\and
School of Physics \& Astronomy, University of Birmingham, Edgbaston, Birmingham, B15 2TT, UK
\and
Leiden Observatory, Leiden University, Postbus 9513, 2300 RA Leiden, Netherlands}

\date{Received; accepted }
\authorrunning{Rickman et al.}
\titlerunning{Three new massive planets and two low mass brown dwarfs at separation larger than 5~AU}


\abstract{
Since 1998, a planet-search around main sequence stars within 50~pc in the southern hemisphere has been carried out with the CORALIE spectrograph at La Silla Observatory.
}{
With an observing time span of more than 20 years, the CORALIE survey is able to detect long term trends in data with masses and separations large enough to select ideal targets for direct imaging. Detecting these giant companion candidates will allow us to start bridging the gap between radial velocity detected exoplanets and directly imaged planets and brown dwarfs.
}{
Long-term precise Doppler measurements with the CORALIE spectrograph reveal radial velocity signatures of massive planetary companions and brown dwarfs on long-period orbits.
}{
In this paper we report the discovery of new companions orbiting HD~181234, HD~13724, HD~25015, HD~92987 and HD~50499. We also report updated orbital parameters for HD~50499b, HD~92788b and HD~98649b. In addition, we confirm the recent detection of HD~92788c. The newly reported companions span a period range of 15.6 to 40.4 years and a mass domain of 2.93 to 26.77 $M_{\mathrm{Jup}}$, the latter of which straddles the nominal boundary between planets and brown dwarfs.
}{
We have reported the detection of five new companions and updated parameters of four known extrasolar planets. We identify at least some of these companions to be promising candidates for imaging and further characterisation.}

\begin{document}
\maketitle
\section{Introduction} \label{sec:1}

Little is known about massive giant planets and brown dwarfs at orbital separations between 5 and 50~AU due to their low occurrence rate \citep{2016PASP..128j2001B} and to the lower sensitivity of the different observing methods in this separation range. Indeed, radial velocities and transit techniques are extremely efficient to detect planets around older stars at short separations \citep{2014prpl.conf..715F}. On the other hand, direct imaging is most efficient at detecting younger planets at separations larger than several times the diffraction limit of the telescope (typically 5 to 10~$\lambda/D$). This translates into several tens of astronomical units for the closest young stellar associations (e.g. $\beta$~Pic and 51~Eri as part of the $\beta$~Pic moving group \citep{2001ApJ...562L..87Z, 2006AJ....131.1730F} and HR~8799 as part of the Columba association \citep{2011ApJ...732...61Z}). And yet, the population of massive giant exoplanets at intermediate orbital separations between 5 - 50~AU is an important puzzle piece needed for constraining the uncertainties that exist in planet formation and evolution models.

The historical \textsc{CORALIE} planet-search survey has been ongoing for more than 20 years in the southern hemisphere and monitors a volume limited sample of 1647 main sequence stars from F8 down to K0 located within 50~pc of the Sun \citep{2000A&A...356..590U}. With an individual measurement precision ranging between 3.5 and 6 $\mathrm{ms}^{-1}$, CORALIE has allowed the detection (or has contributed to the detection) of more than 140 extra-solar planet candidates \citep{2002A&A...388..632P, 2002A&A...390..267U, 2008A&A...480L..33T, 2010A&A...511A..45S, 2013A&A...551A..90M}. Such a long and continuous monitoring of nearby main sequence stars is unique among all planet search surveys. It allows us to detect massive giant planets at separations larger 5~AU as well as to identify small radial velocity drifts hinting for the presence of low mass companions at even wider separations.

These are indeed golden targets for direct imaging, as such old and very low mass companions are rare and very difficult to search for blindly. \cite{2018A&A...614A..16C} has shown with the discovery of the ultra cool brown dwarf companion orbiting the planet host star HD~4113~A that long term radial velocity surveys are an extremely useful tool to select targets to image.  Not only does it allow us to start filling in a largely unexplored parameter space, but through combining radial velocity and direct imaging we can now expect to measure the masses of these companions using Kepler's laws. By constraining the mass, we are able to place additional constraints on the evolution of the companion, both in terms of temperature and atmospheric composition \citep{2018ApJ...853..192C,2018arXiv180505645P}.

In this paper we report the discovery of four new giant planets and brown dwarfs orbiting HD~181234, HD~13724, HD~25015 and HD~92987, together with the updated CORALIE orbital elements for an already known exoplanet around  HD~98649 \citep{2013A&A...551A..90M}. We also report updated orbital parameters for HD~50499b \citep{2005ApJ...632..638V}, as well as the detection of HD 50499c, which has previously been noted by \cite{2005ApJ...632..638V}, \cite{2017AJ....153..208B} and \cite{2018arXiv180408329B}. We also report the updated orbital parameters of HD~92788b detected by \cite{2001ApJ...551.1107F} and confirm the recent detection of HD~92788c \citep{2019MNRAS.484.5859W}.

The paper is organised as follows. The host stars' properties are summarised in Sect.~\ref{sec:2}. In Section~\ref{sec:3} we present our radial velocity data and the inferred orbital solution of the newly detected companions. In Section~\ref{sec:4} we present the CORALIE updated parameters of already known exoplanets with new detections in two of these systems. The results are discussed in Sect.~\ref{sec:5} with some concluding remarks.

\section{Stellar Characteristics} \label{sec:2}

\begin{table*}
\caption{Observed and inferred stellar parameters for host stars to the planet candidates - HD~181234, HD~25015, HD~50499 and HD~92788, HD~98649.}
\label{table:1}      % is used to refer this table in the text
\centering           % used for centering table
\begin{tabular}{c c c c c c c c c c}        % centered columns (4 columns)
\hline\hline                    % inserts double horizontal lines
Parameters & Units & HD~181234 & HD~25015 & HD~50499 & HD~92788 & HD~98649 \\    % table heading
\hline    
Spectral Type\tablefoottext{a} & & G5 & K1V & G1V & G6V & G3/G5V \\
$V$\tablefoottext{a} & & 8.59 & 8.87 & 7.21 & 7.31 & 8.00 \\
$B-V$\tablefoottext{a} & & 0.841 & 0.899 & 0.614 & 0.694 & 0.658 \\ 
 $\pi$\tablefoottext{b} & $[\text{mas}]$ & $20.9 \pm 0.06$ & $26.7 \pm 0.05$ & $21.58 \pm 0.03$ & $28.83 \pm 0.05$ & $23.7 \pm 0.05$ \\
 $L$\tablefoottext{b} & $[\text{L}_{\odot}]$ & $0.80 \pm 0.003$ & $0.41 \pm 0.001$ & $2.38 \pm 0.005$ & $1.25 \pm 0.003$ & $0.98 \pm 0.003$ \\
 $T_{\text{eff}}$\tablefoottext{c} & $[\text{K}]$ & $5386 \pm 60$ & $5160 \pm 63$ &  $6099 \pm 43$ & $5744 \pm 24$ \tablefoottext{e} & $5790 \pm 58$ \\
 $\log g$\tablefoottext{c} & $[\text{cgs}]$ & $4.25 \pm 0.11$ & $4.40 \pm 0.14$ &  $4.42 \pm 0.05$ & $4.39 \pm 0.04$ & $4.51 \pm 0.09$ \\
 $[\text{Fe/H}]$\tablefoottext{c} & $[\text{dex}]$ & $0.32 \pm 0.05$ & $0.04 \pm 0.04$ & $0.38 \pm 0.03$ & $0.27 \pm 0.02$ \tablefoottext{e} & $0.05 \pm 0.04$ \\
  $\log R^{'}_{HK}$\tablefoottext{c} & & $-5.17 \pm 0.01$ & $-4.48 \pm 0.002$ &  $-5.08 \pm 0.004$ & $-4.98 \pm 0.01$  \tablefoottext{f} & $-5.06 \pm 0.005$\\
$P_{\text{rot}}$ & $[\text{days}]$ & $50.8 \pm 2.0$ & $13.6 \pm 2.3$ &  $22.4\pm1.0$ & $31.0\pm1.4$ \tablefoottext{g} & $27.7\pm1.2$ \\
$v \sin i$\tablefoottext{d} & $[\text{km s}^{-1}]$ & 2.105 & 3.485 & 4.313 & 2.719 & 2.218 \\
\hline
$M_{*}$ & $[\text{M}_{\odot}]$ & $1.01 \pm 0.06$ & $0.86 \pm 0.05$ & $1.31 \pm 0.07$ & $1.15 \pm 0.07$ & $1.03 \pm 0.06$ \\
$R_{*}$ & $[\text{R}_{\odot}]$ & $1.05 \pm 0.07$ & $0.83 \pm 0.04$ & $1.42 \pm 0.02$ &  $1.14 \pm 0.02$ & $1.01 \pm 0.02$ \\
Age & $[\text{Gyr}]$ & $6.32 \pm 2.58$ & $4.00 \pm 3.41$ & $2.40 \pm 0.56$ &  $2.55 \pm 1.51$ & $2.42 \pm 1.62$ \\
\hline
\end{tabular}
\tablebib{(1)~\citet{2018arXiv180409365G}; (2)~\citet{2008ApJ...687.1264M}; (3)~\citet{1997A&A...323L..49P};  (4)~\citet{2008A&A...487..373S}}
\tablefoot{
\tablefoottext{a}{Parameters taken from HIPPARCOS \citep{1997A&A...323L..49P}}
\tablefoottext{b}{Parameters taken from \emph{Gaia} data release 2 \citep{2018arXiv180409365G}}
\tablefoottext{c}{Parameters derived using CORALIE spectra.}
\tablefoottext{d}{Parameters derived using CORALIE CCF.}
\tablefoottext{e}{Parameters taken from \cite{2008A&A...487..373S}.}
\tablefoottext{f}{Parameters derived using HARPS spectra.}
\tablefoottext{g}{From the calibration of the rotational period vs. activity \citep{2008ApJ...687.1264M}.}}
\end{table*}

\begin{table*}
\caption{Observed and inferred stellar parameters for host stars to the brown dwarf candidates - HD~13724 and HD~92987.}
\label{table:2}      % is used to refer this table in the text
\centering           % used for centering table
\begin{tabular}{c c c c c c}        % centered columns (4 columns)
\hline\hline                    % inserts double horizontal lines
Parameters & Units & HD~13724 & HD~92987 \\    % table heading
\hline                          % inserts single horizontal line
    Spectral Type\tablefoottext{a}& & G3/G5V & G2/G3V \\ %inserting body of the table
    $V$\tablefoottext{a} & & 7.89 & 7.03 \\
    $B-V$\tablefoottext{a} & & 0.667 & 0.641 \\
    $\pi$\tablefoottext{b}& $[\text{mas}]$ & $23.0 \pm 0.03$ & $22.9 \pm 0.03$ \\
    $L$\tablefoottext{b} & $[\text{L}_{\odot}]$ & $1.14^{+0.001}_{-0.002}$ & $2.55 \pm 0.006$ \\
    $T_{\text{eff}}$\tablefoottext{c} & $[\text{K}]$ & $5868\pm27$\tablefoottext{e} & $5770\pm36$\tablefoottext{f} \\
    $\log g$\tablefoottext{c} & $[\text{cgs}]$ & $4.44 \pm 0.07$\tablefoottext{g} & $4.00\pm0.15$\tablefoottext{f} \\
    $[\text{Fe/H}]$\tablefoottext{c} & $[\text{dex}]$ & $0.23\pm0.02$\tablefoottext{e} & $-0.08\pm0.08$\tablefoottext{f} \\
    $\log R^{'}_{HK}$\tablefoottext{c} & & $-4.76 \pm .003$ & $-5.090 \pm 0.006 $ \\
    $P_{\text{rot}}$ & $[\text{days}]$ & $20.2\pm1.2$ & $26.2\pm1.1$ \\
    $v \sin i$\tablefoottext{d} & $[\text{km s}^{-1}]$ & 3.025 & 2.616 \\
    \hline
    $M_{*}$ & $[\text{M}_{\odot}]$ & $1.14 \pm 0.06$ & $1.08 \pm 0.06$ \\
    $R_{*}$ & $[\text{R}_{\odot}]$ & $1.07 \pm 0.02$ & $1.58 \pm 0.04$ \\
    Age & $[\text{Gyr}]$ & $0.76 \pm 0.71$ & $7.75 \pm 0.31$ \\ 
\hline                              %inserts single line
\end{tabular}
\tablebib{(1)~\citet{2006MNRAS.370..163B}; ; (2)~\citet{2018arXiv180409365G}; (3)~\citet{2014A&A...566A..66G}; (4)~\citet{1997A&A...323L..49P}; (5)~\citet{2014A&A...563A..52P}}
\tablefoot{
\tablefoottext{a}{Parameters taken from HIPPARCOS \citep{1997A&A...323L..49P}.}
\tablefoottext{b}{Parameters taken from \emph{Gaia} data release 2 \citep{2018arXiv180409365G}.}
\tablefoottext{c}{Parameters derived using CORALIE spectra.}
\tablefoottext{d}{Parameters derived using CORALIE CCF.}}
\tablefoottext{e}{Parameters taken from \cite{2014A&A...566A..66G}.}
\tablefoottext{f}{Parameters taken from \cite{2006MNRAS.370..163B}.}
\tablefoottext{g}{Parameters taken from \cite{2014A&A...563A..52P}.}
\end{table*}

Spectral types, V band magnitude and colour indices are taken from the HIPPARCOS catalogue \citep{1997A&A...323L..49P} while astrometric parallaxes ($\pi$) and luminosities are taken from the second \emph{Gaia} date release \citep{2018arXiv180409365G}. Effective temperatures, gravities and metallicities are derived using the same spectroscopic methods as applied in \cite{2013A&A...556A.150S}, whilst the $ v \sin (i) $ is computed using the calibration of CORALIE's Cross Correlation Function (CCF) \citep{2001A&A...373.1019S, marmier_phd_thesis}. 

The mean chromospheric activity index - $\log{(R^{'}_{HK})}$ - of each star is computed by co-adding the corresponding CORALIE spectra to improve the  signal-to-noise which allows us to measure the Ca II re-emission at $\lambda = 3933.66$~\AA. We derived an estimate of the rotational period  the star from the  mean  $\log{(R^{'}_{HK})}$ activity index using the calibration of \cite{2008ApJ...687.1264M}.

Stellar radii and their uncertainties are derived from the \emph{Gaia} luminosities and the effective temperatures obtained from the spectroscopic analysis. A systematic error of 50~K was quadratically added to the effective temperature error bars and was propagated in the radius uncertainties.

The mass and the age of the stars, as well as their uncertainties, are derived using the Geneva stellar evolution models \citep{2012A&A...537A.146E, 2013A&A...558A.103G}. The interpolation in the model grid was made through a Bayesian formalism using observational Gaussian priors on $T_{\mathrm{eff}}$, $M_{V}$, $\log g$, and [Fe/H] \citep{marmier_phd_thesis}. 

The observed and inferred stellar parameters for newly detected host stars to planetary companions are summarised in Table~\ref{table:1} and host stars to brown dwarf companions in Table~\ref{table:2}. 

\section{Radial velocities and orbital solutions} \label{sec:3}

\begin{table*}
\caption{Best-fitted solutions for the substellar companions orbiting HD~13724, HD~181234, HD~25015 and HD~92987. For each parameter, the mode of the posterior is considered, with error bars computed from the MCMC chains with 10,000,000 iterations using a 68.27\% confidence interval.}     % title of Table
\label{table:3}      % is used to refer this table in the text
\centering           % used for centering table
\begin{tabular}{c c c c c c c c }        % centered columns (4 columns)
\hline\hline                    % inserts double horizontal lines
Parameters & Units & HD~13724 & HD~181234 & HD~25015 & HD~92987 \\    % table heading
\hline     % inserts single horizontal line
    $P$ & $\text{[years]}$ & $40.42^{+13.42}_{-4.38}$ & $20.43^{+0.22}_{-0.21}$ & $16.48^{+1.86}_{-0.72}$ & $28.35^{+1.51}_{-0.74}$ &  \\
    $K$ & $\text{[ms}^{-1}\text{]}$ & $214.3^{+21.5}_{-10.2}$ & $126.8^{+1.8}_{-1.6}$ & $60.1^{+3.1}_{-3.2}$ & $152.7^{+2.3}_{-2.7}$ \\
    $e$ & & $0.34^{+0.09}_{-0.05}$ & $0.73 \pm 0.01$ & $0.39^{+0.09}_{-0.07}$ & $0.21^{+0.02}_{-0.01}$ \\
    $\omega$ & $[\text{deg}]$ & $187.5^{+2.9}_{-1.7}$ & $93.3^{+1.7}_{-1.8}$ & $77.7^{+9.8}_{-8.6}$ & $195.1^{+6.7}_{-8.4}$ \\
    $T_p$ & $[\text{JD}]$ & $6189.1^{+73}_{-54}$ & $7668.7^{+5.5}_{-5.0}$ & $5852^{+160}_{-140}$ & $7889^{+130}_{-180}$ \\
    \hline
    $M.\sin i$ & $[M_{\text{Jup}}]$ & $26.77^{+4.4}_{-2.2}$ & $8.37^{+0.34}_{-0.36}$ &  $4.48^{+0.30}_{-0.28}$ & $16.88^{+0.69}_{-0.65}$ \\
    $a$ & $[\text{AU}]$ & $12.40^{+2.6}_{-0.9}$ & $7.52^{+0.16}_{-0.16}$ & $6.19^{+0.45}_{-0.23}$ & $9.62^{+0.36}_{-0.26}$ \\
    \hline
    $N_{\text{RV}}$ & & 167 & 115 & 110 & 93 \\
    $\Delta T$ & $[\text{years}]$ & 19.3 & 18.6 & 17.6 & 19.9 \\
\hline                              %inserts single line
\end{tabular}
\tablefoot{
$\Delta T$ is the time interval between the first and last measurements. C98 stands for CORALIE-98, C07 for CORALIE-07 and C14 for CORALIE-14. $N_{\text{RV}}$ is the number of RV measurements. $T_P$ is shown in BJD-2,450,000.}
\end{table*}

The CORALIE observations span over more than 20 years, from June 1998 to December 2018. During that time, CORALIE went through two major upgrades, in June 2007 \citep{2010A&A...511A..45S} and in November 2014 to increase overall efficiency and accuracy of the instrument. These changes introduced small offsets in the measured radial velocities that depend on several parameters such as the spectral type of the star and its systemic velocity. For this reason, we decided to consider CORALIE as three different instruments, corresponding to the different upgrades: the original CORALIE as CORALIE-98 (C98), the first upgrade as CORALIE-07 (C07) and the latest upgrade as CORALIE-14 (C14). 

In addition to the radial velocity time series, the CORALIE automated pipeline also provides several useful indicators that help pinpoint the origin of observed periodic signals. These are the CCF Full Width at Half Maximum (FWHM), bisector and the $H_\alpha$ chromospheric activity indicator.

We are also using published radial velocities taken with other spectrographs, namely, HARPS \citep{2003Msngr.114...20M}, HIRES \citep{1994SPIE.2198..362V} and HAMILTON \citep{1538-3873-99-621-1214}. The data products presented in this paper are available at the Data and Analysis Center for Exoplanets (\emph{DACE})\footnote{The data are available at `The Data and Analysis Center for Exoplanets' (\emph{DACE}) which can be accessed at \url{https://dace.unige.ch}.}.

We perform an initial modeling of the radial velocity time series using the online \emph{DACE} platform. Keplerian model initial conditions are computed using the formalism described in \cite{2016A&A...590A.134D}. The stellar activity detrending and the modeling of the instrumental noise and the stellar jitter follow the formalism described in \cite{2016A&A...585A.134D} and \cite{2018A&A...614A.133D}. Analytical false alarm probabilities (FAP) are computed on the periodogram of the residuals following \cite{2008MNRAS.385.1279B} and numerical FAP values which are used in this paper are computed by permutation of the calendar. Periodic signals with a FAP lower than 0.1\% are considered significant and are added to the model. For each periodogram shown, the three lines represent the $10 \%$, $1 \%$ and $0.1 \%$ false alarm probability in ascending order. 

Once the RV time series is fully modeled using \emph{DACE} online tools, we run a Markov Chain Monte Carlo (MCMC) analysis of each system using the algorithm described in \cite{2014MNRAS.441..983D,2016A&A...585A.134D} and \cite{2018A&A...614A.133D} to obtain the posterior distributions of the model parameters. Each MCMC simulation is run with 10,000,000 iterations drawing the proposal solution obtained using \emph{DACE}.
 The parameter confidence intervals are computed for a $68.27 \%$ confidence level.

Gaussian priors are set for the instrument offsets and stellar mass with uniform priors for the orbital elements. In the cases where the minimum radial velocity (${T}_{{V}_{\mathrm{min}}}$) or the maximum radial velocity (${T}_{{V}_{\mathrm{max}}}$) is well sampled, we perform the fit using either of these instead of fitting the phase.

In this section we present the orbital solutions for newly reported giant planets and brown dwarfs from the CORALIE survey. A summary of the orbital solutions can be found in Table~\ref{table:3} for the newly detected companions and the fully probed MCMC parameter space is shown in the Appendix.

\subsection{HD~181234 (LTT~5654, HIP~95015)}

\begin{figure}
  \resizebox{\hsize}{!}{\includegraphics{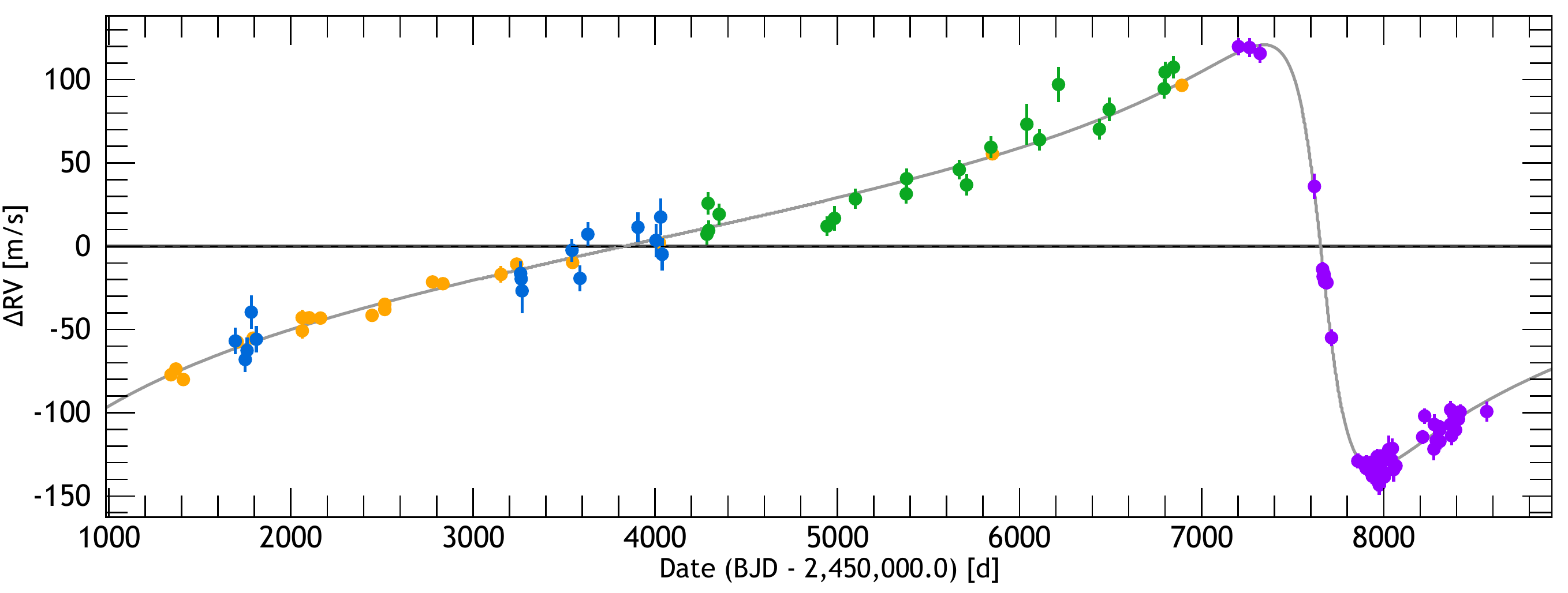}}
  \resizebox{\hsize}{!}{\includegraphics{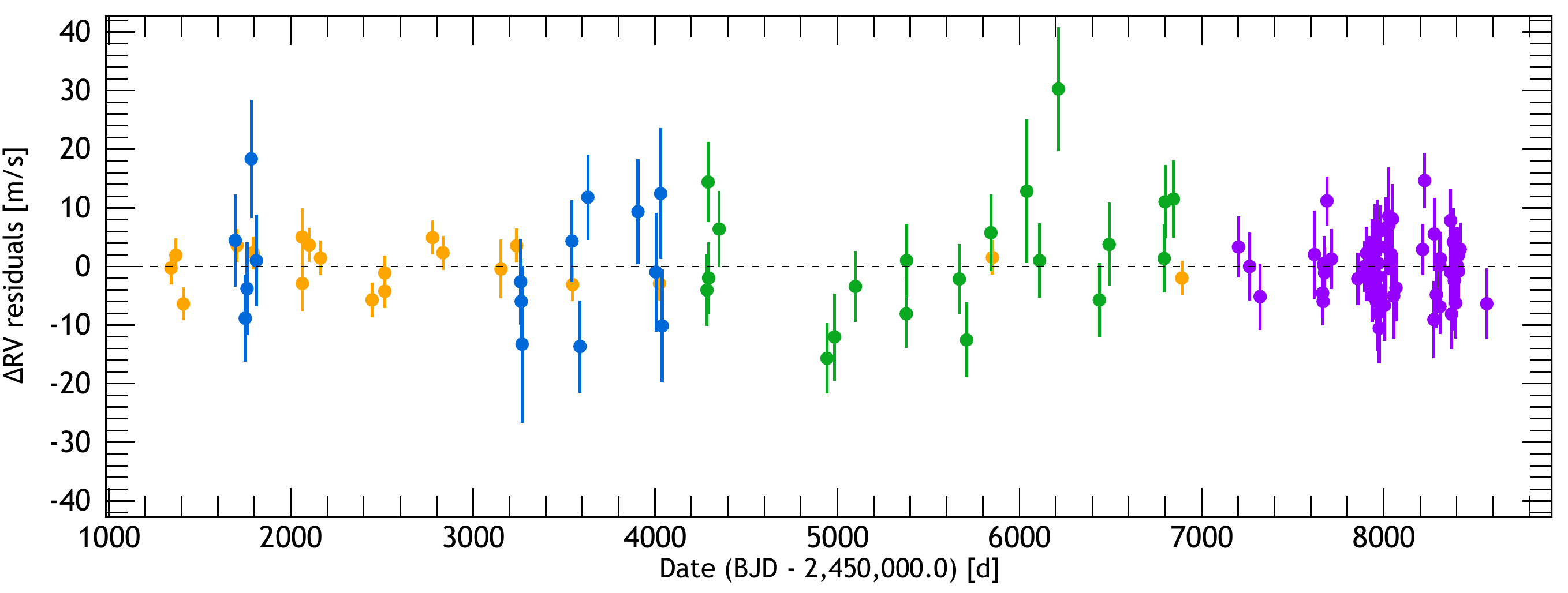}}
  \resizebox{\hsize}{!}{\includegraphics{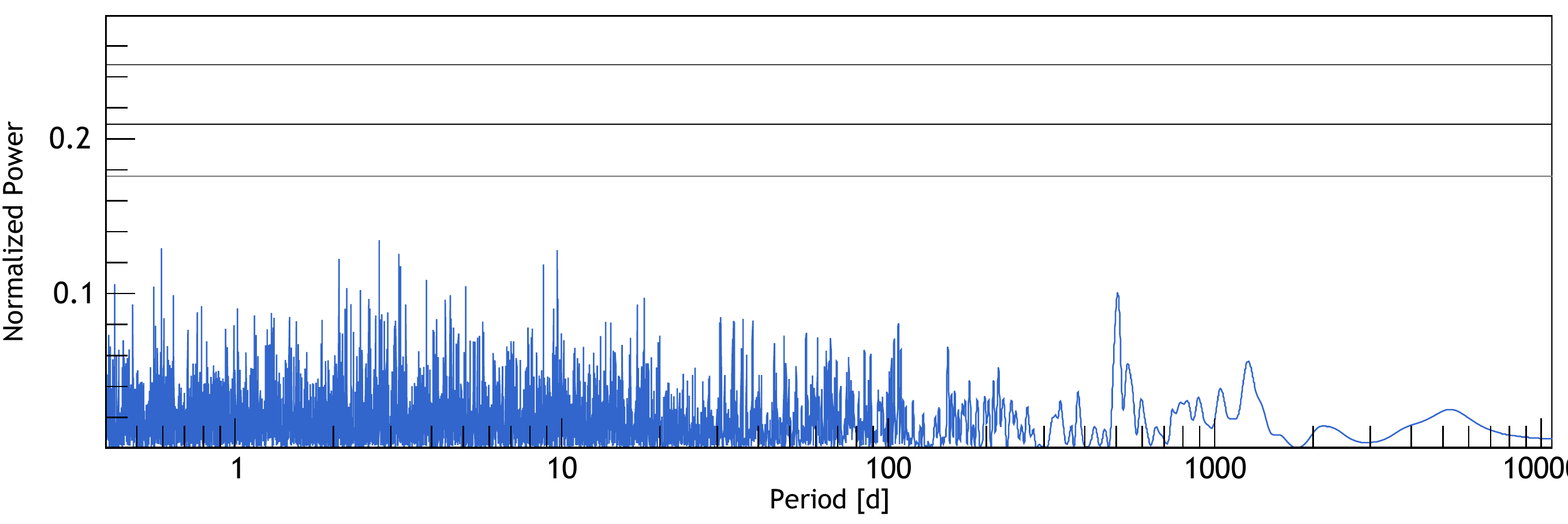}}
  \caption{\textbf{Top:} HD~181234 Radial velocity measurements as a function of Julian Date obtained with CORALIE-98 (blue), CORALIE-07 (green), CORALIE-14 (purple) and HIRES data \citep{2017AJ....153..208B} in orange. The best single-planet Keplerian model is represented as a black curve. \textbf{Middle:} The RV residuals of HD~181234. \textbf{Bottom:} The periodogram of the residuals for HD~181234. The three black lines represent the $10 \%$, $1 \%$ and $0.1 \%$ false alarm probability in ascending order.}
  \label{fig:1}
\end{figure}

HD~181234 has been observed with CORALIE at La Silla Observatory since May 2000. Fifteen measurements were taken with CORALIE-98, 21 additional radial velocity measurements were obtained with CORALIE-07 and 59 additional radial velocity measurements were obtained with CORALIE-14. HD~181234 has also been observed with Keck/HIRES \citep{2017AJ....153..208B} with 20 radial velocity measurements from June 1999 to August 2014.

The best fit Keplerian, as shown in Fig.~\ref{fig:1}, shows that we are looking at a highly eccentric system with an eccentricity of 0.73. It has an orbital period of 20.4 years with a minimum mass of 8.4 $M_{\mathrm{Jup}}$. The orbital solutions are summarised in Table~\ref{table:3}. Figure~\ref{fig:1} shows the CORALIE radial velocities and the corresponding best-fit Keplerian model along with the radial velocity residuals and a periodogram of the residuals. The results from the fully probed parameter space from the MCMC are shown in the appendix.

\subsection{HD~92987 (HIP~52472)}

\begin{figure}
  \resizebox{\hsize}{!}{\includegraphics{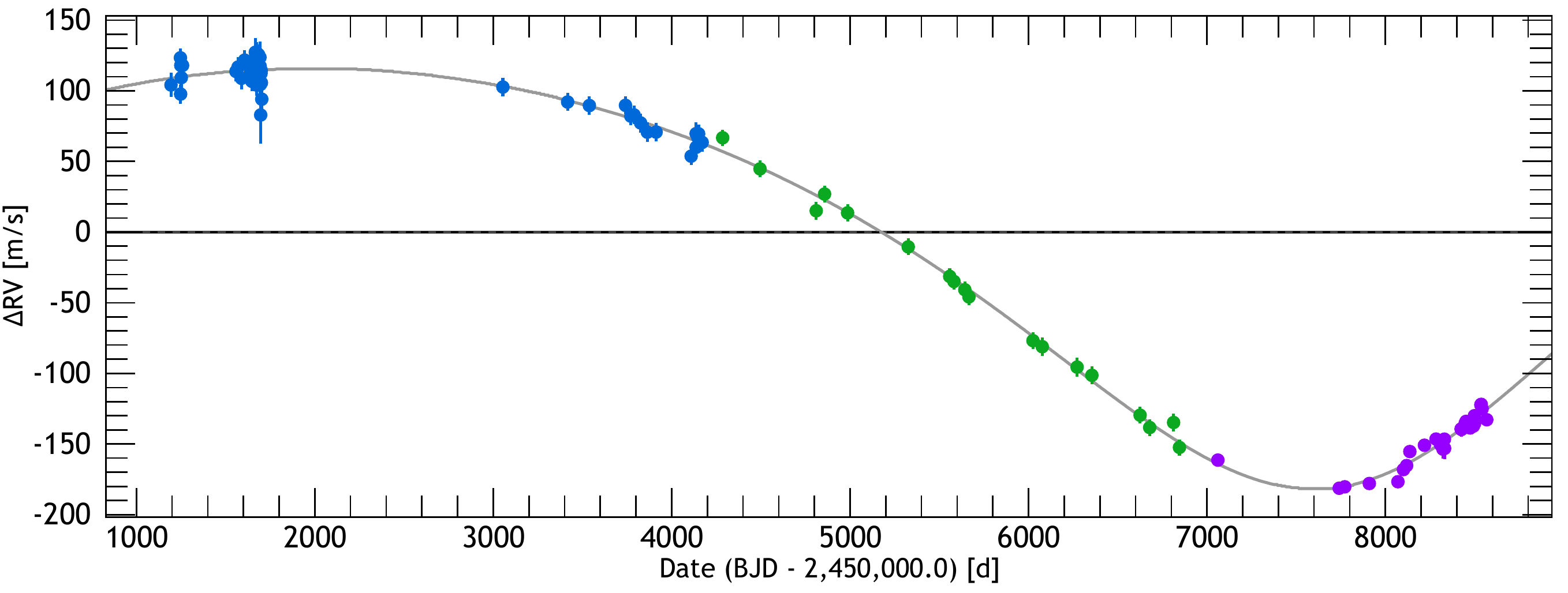}}
  \resizebox{\hsize}{!}{\includegraphics{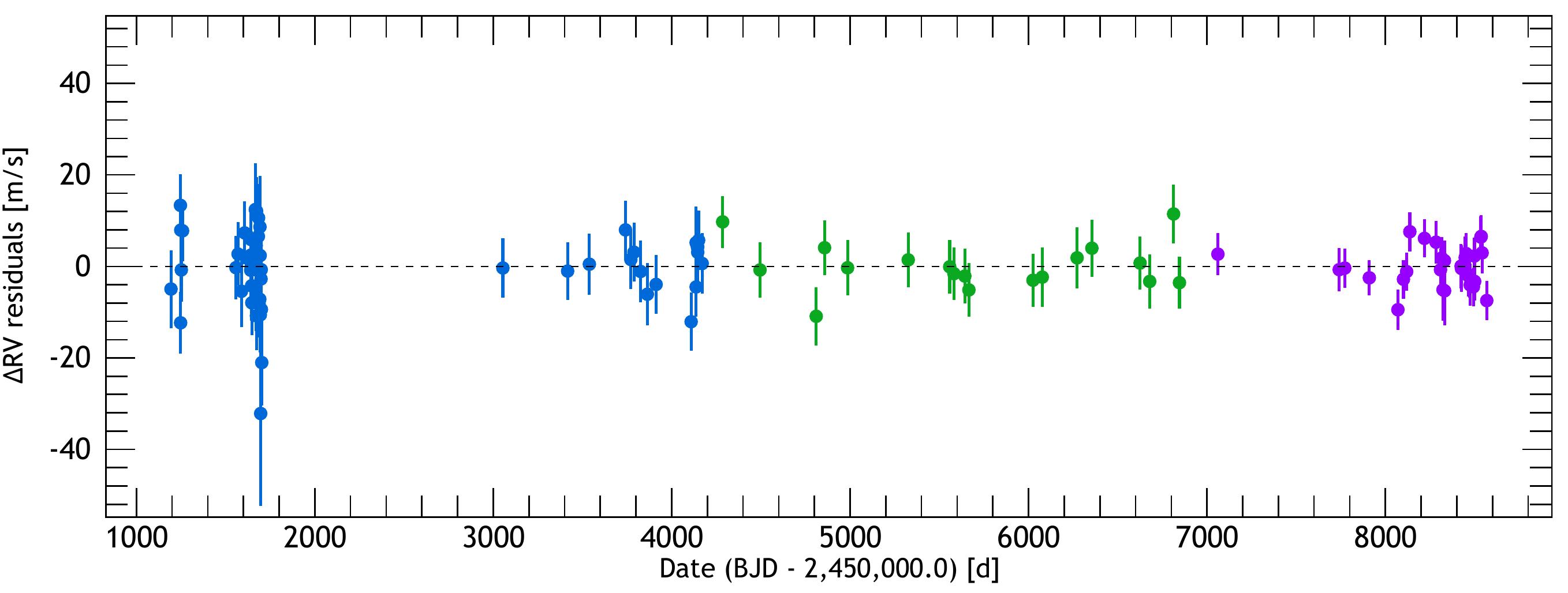}}
    \resizebox{\hsize}{!}{\includegraphics{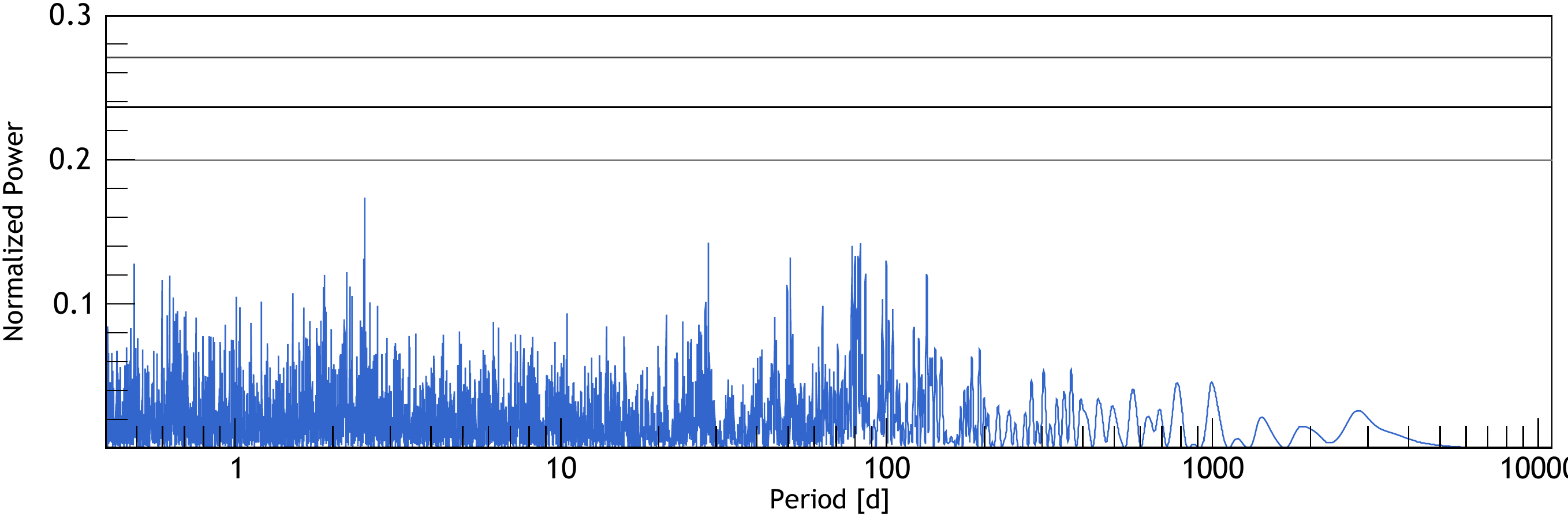}}
  \caption{\textbf{Top:} HD~92987 Radial velocity measurements as a function of Julian Date obtained with CORALIE-98 (blue), CORALIE-07 (green) and CORALIE-14 (purple). The fitted single-planet Keplerian model is represented as a black curve. \textbf{Middle:} The RV residuals of HD~92987. \textbf{Bottom:} The periodogram of the residuals for HD~92987. The three black lines represent the $10 \%$, $1 \%$ and $0.1 \%$ false alarm probability in ascending order.}
  \label{fig:2}
\end{figure}

HD~92987 has been observed with CORALIE at La Silla Observatory since January 1999. Fifty-three measurements were taken with CORALIE-98, 18 additional radial velocity measurements were obtained with CORALIE-07 and 22 additional radial velocity measurements were obtained with CORALIE-14.

HD~92987 is one of the brown dwarf candidates with a minimum mass of 16.88 $M_{\mathrm{Jup}}$ and a semi-major axis of 9.62 AU, making it a promising candidate for direct imaging. The orbital solutions for HD~92987 are summarised in Table~\ref{table:3}. Figure~\ref{fig:2} shows the CORALIE radial velocities and the corresponding best-fit Keplerian model along with the radial velocity residuals and a periodogram of the residuals. The results from the fully probed parameter space from the MCMC are shown in the appendix.

\subsection{HD~25015 (HIP~18527)}

\begin{figure}
  \resizebox{\hsize}{!}{\includegraphics{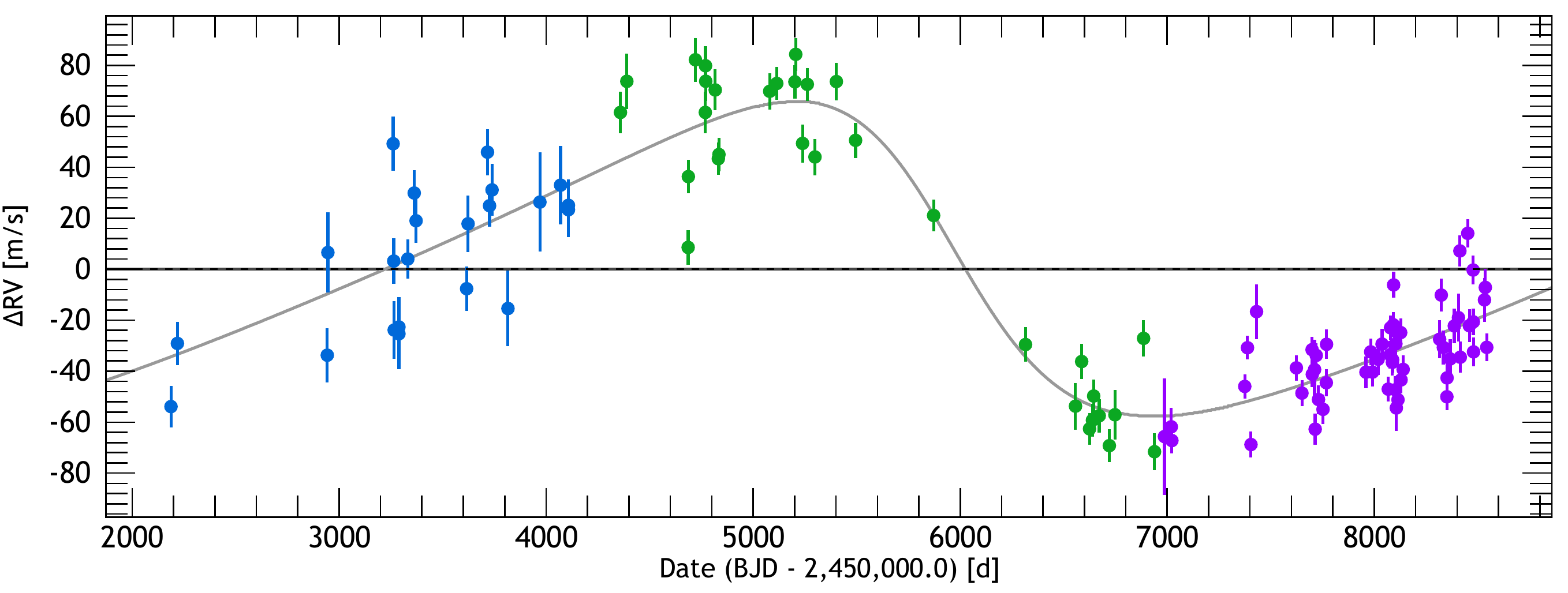}}
  \resizebox{\hsize}{!}{\includegraphics{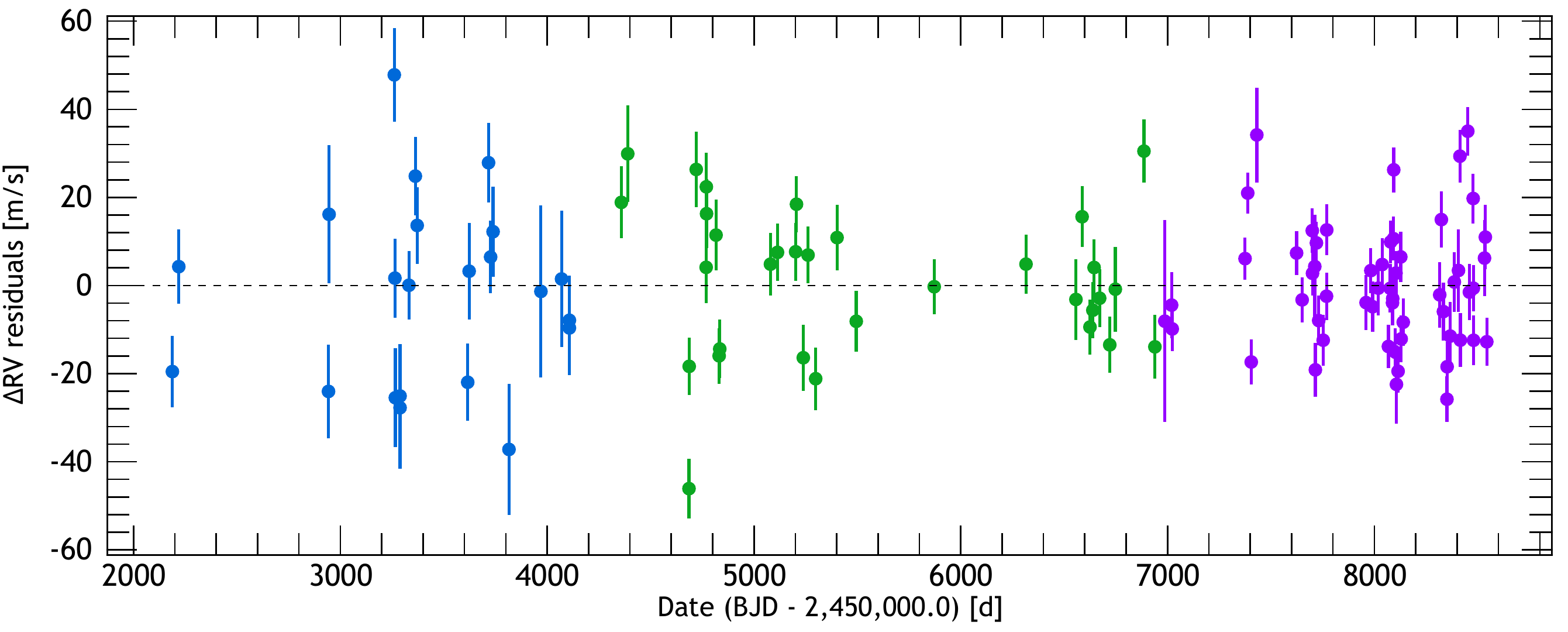}}
  \resizebox{\hsize}{!}{\includegraphics{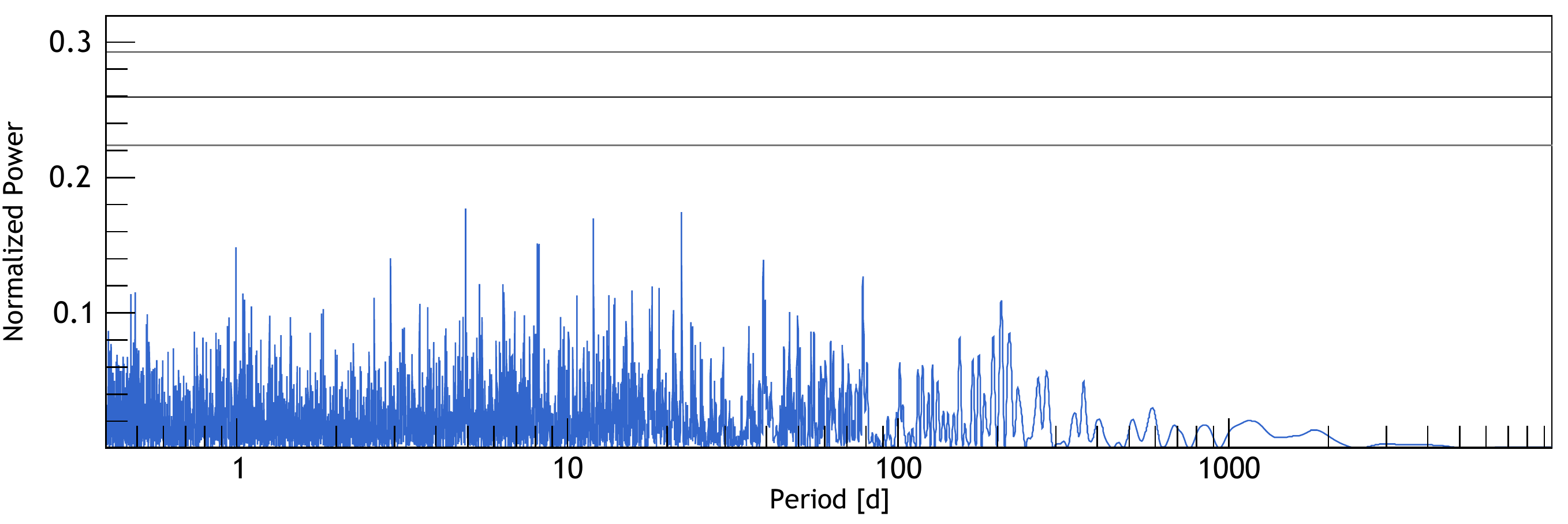}}
   \resizebox{\hsize}{!}{\includegraphics{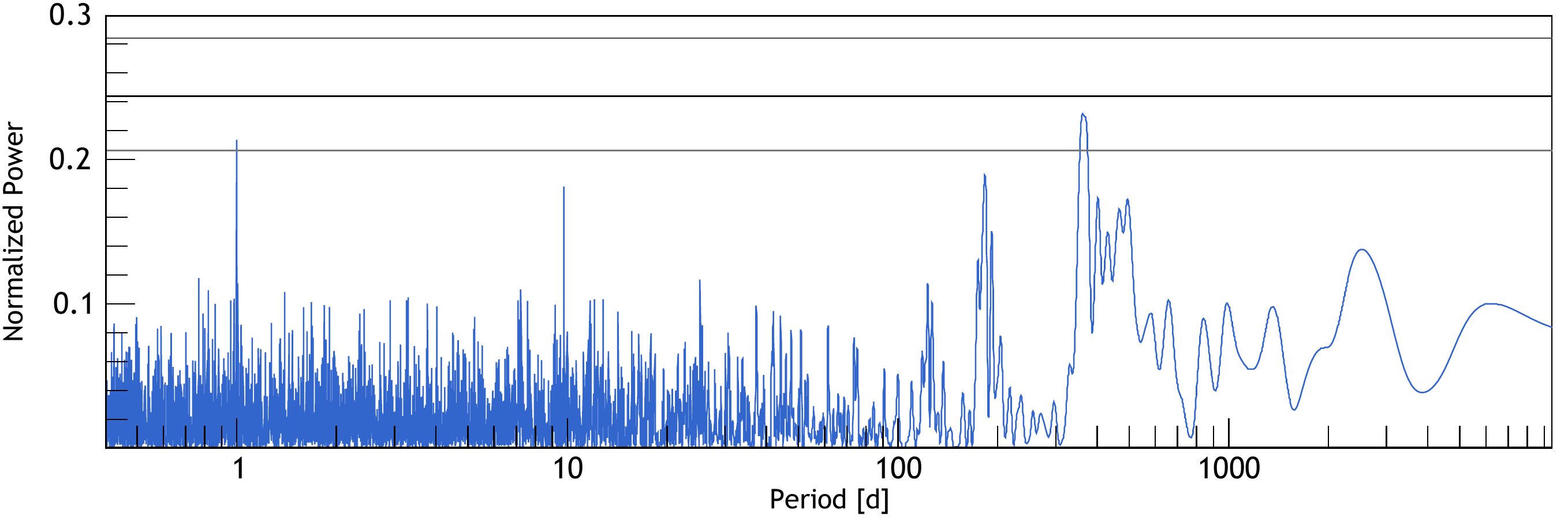}}
  \caption{\textbf{Top:} HD~25015 Radial velocity measurements as a function of Julian Date obtained with CORALIE-98 (blue), CORALIE-07 (green) and CORALIE-14 (purple). The fitted single-planet Keplerian model is represented as a black curve. \textbf{Second figure:} The RV residuals of HD~25015. \textbf{Third figure:} The periodogram of the residuals for HD~25015 after the signal has been removed showing no significant signals. The three black lines represent the $10 \%$, $1 \%$ and $0.1 \%$ false alarm probability in ascending order. \textbf{Bottom:} Periodogram of $H_\alpha$ before detrending. The three black lines represent the $10 \%$, $1 \%$ and $0.1 \%$ false alarm probability in ascending order, showing a significant peak above the $10 \%$ FAP at 370.26 days.}
  \label{fig:3}
\end{figure}

HD~25015 has been observed with CORALIE at La Silla Observatory since May 2001. Twenty-two measurements were taken with CORALIE-98, 32 additional radial velocity measurements were obtained with CORALIE-07, followed by 56 additional radial velocity measurements obtained with CORALIE-14.

We note here that all of the stars in our sample are very quiet with the exception of HD~25015. When we plot the $H_\alpha$ versus time, we see that some periodicity exists in the periodogram at a false alarm probability of greater than $10\%$, as seen in Fig~\ref{fig:3} at approximately 370.26 days. We do not see a periodicity in the FWHM or bisector, therefore we cannot exclude that there is telluric line contamination in the $H_\alpha$ index with CORALIE. We detrend the radial velocity from the stellar activity using the $H_\alpha$ indicators. A scale factor is adjusted to the smoothed $H_\alpha$ (using a Gaussian filter at 0.1 years), as well as an additional jitter term proportional to the activity trend.

The orbital solutions for HD~25015 are summarised in Table~\ref{table:3}. Figure~\ref{fig:3} shows the CORALIE radial velocities and the corresponding best-fit Keplerian model along with the radial velocity residuals, the periodogram of the residuals and a periodogram for $H_\alpha$ before detrending. The results from the fully probed parameter space from the MCMC are shown in the appendix.

\subsection{HD~13724 (HIP~10278)}

\begin{figure}
  \resizebox{\hsize}{!}{\includegraphics{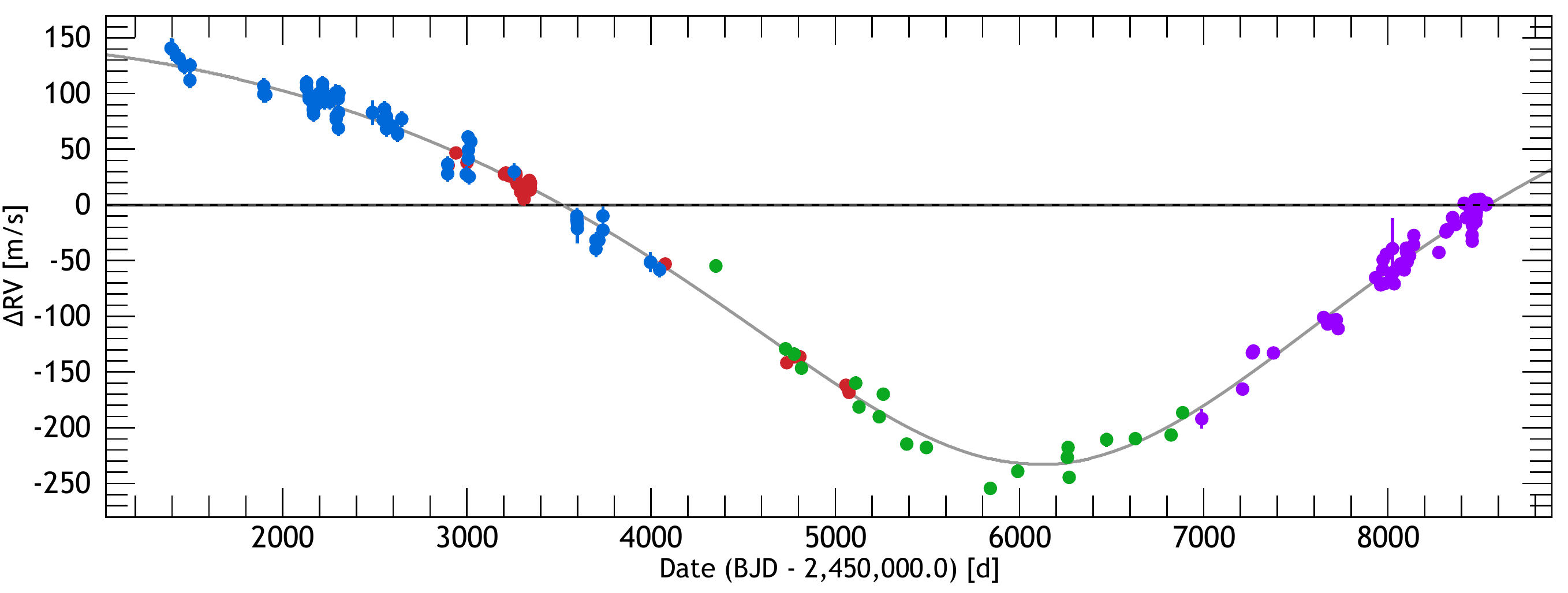}}
  \resizebox{\hsize}{!}{\includegraphics{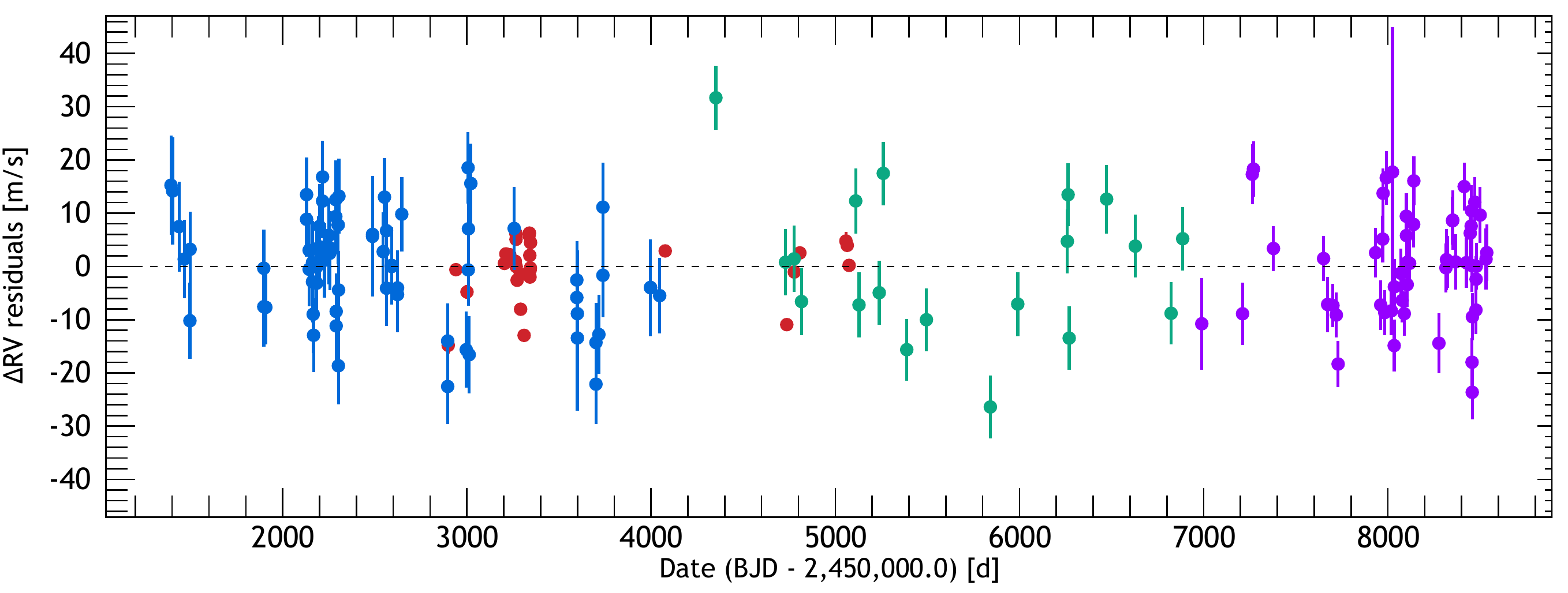}}
    \resizebox{\hsize}{!}{\includegraphics{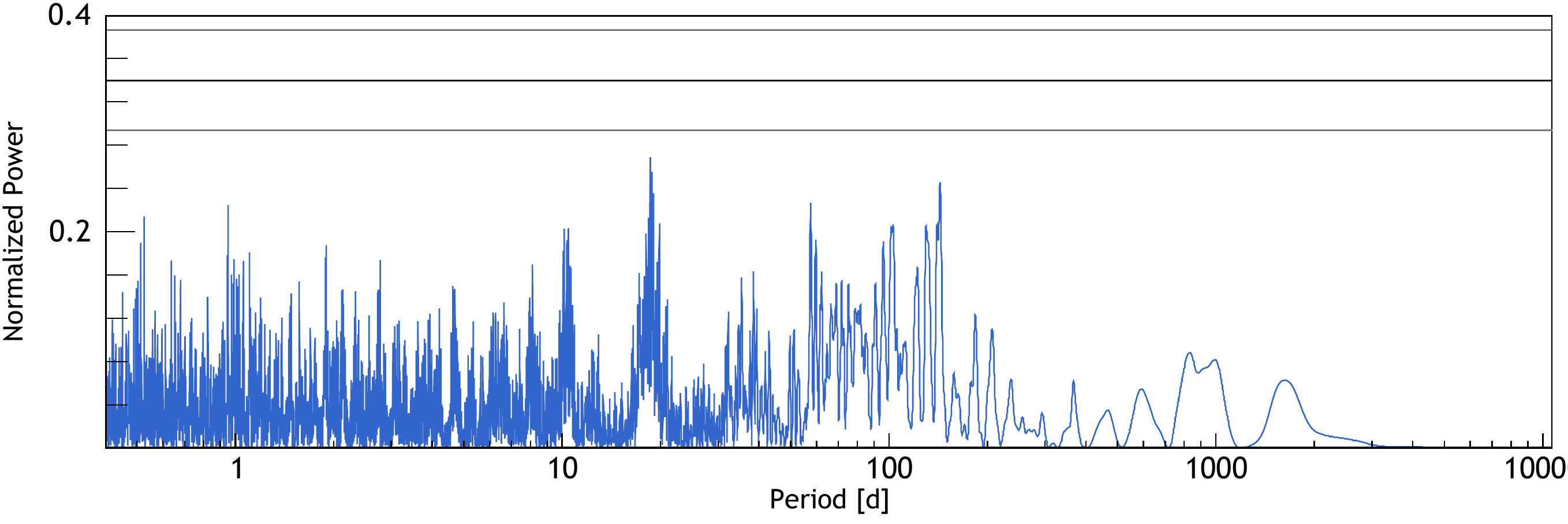}}
  \caption{\textbf{Top:} HD~13724 Radial velocity measurements as a function of Julian Date obtained with CORALIE-98 (blue), CORALIE-07 (green), CORALIE-14 (purple) and HARPS (red). The best single-planet Keplerian model is represented as a black curve. \textbf{Middle:} The RV residuals of HD~13724. \textbf{Bottom:} The periodogram of the residuals for HD~13724 after the signal has been removed showing no significant signals. The three black lines represent the $10 \%$, $1 \%$ and $0.1 \%$ false alarm probability in ascending order.}
  \label{fig:4}
\end{figure}

HD~13724 has been observed with CORALIE at La Silla Observatory since August 1999. It is a relatively young star at $0.76 \pm 0.71$ Gyr old.

During the past 19.3 years, 167 Doppler measurements were taken on this target with 70 radial velocity measurements taken with CORALIE-98, 19 with CORALIE-07, 48 with CORALIE-14 and 30 with HARPS. HD~13724 is a brown dwarf companion with a minimum mass of 26.77 $M_{\mathrm{Jup}}$ and a semi-major axis of 12.4 AU, making it a promising candidate for direct imaging.

The orbital solutions for HD~13724 are summarised in Table~\ref{table:3}. Figure~\ref{fig:4} shows the CORALIE radial velocities and the corresponding best-fit Keplerian model along with the radial velocity residuals and a periodogram of the residuals. The results from the fully probed parameter space from the MCMC are shown in the appendix.
 
\section{Updated parameters for known exoplanets} \label{sec:4}

We present updated orbital parameters for a known exoplanet around HD~98649 \citep{2013A&A...551A..90M}. We also report updated orbital parameters for HD~92788b \citep{2001ApJ...551.1107F} and confirm the detection of HD~92788c \citep{2019MNRAS.484.5859W}. Moreover, we report the discovery of a new planet around HD~50499 and report updated orbital parameters for HD~50499b \citep{2005ApJ...632..638V}.

The radial velocity data is fitted in the same way as described in section~\ref{sec:3}.The stellar parameters for these systems are summarised in Table~\ref{table:1}. For HD~50499 we also report the discovery of a new exoplanet in each system.

The orbital parameters for HD~98649, HD~50499 and HD~92788 are summarised in Table~\ref{table:4}. The probed physical parameters using the MCMC for each target are shown in the Appendix.

\subsection{A 7 Jupiter-mass companion on an eccentric orbit of 15 years around HD~98649 (LTT~4199, HIP~55409)}

\begin{figure}
  \resizebox{\hsize}{!}{\includegraphics{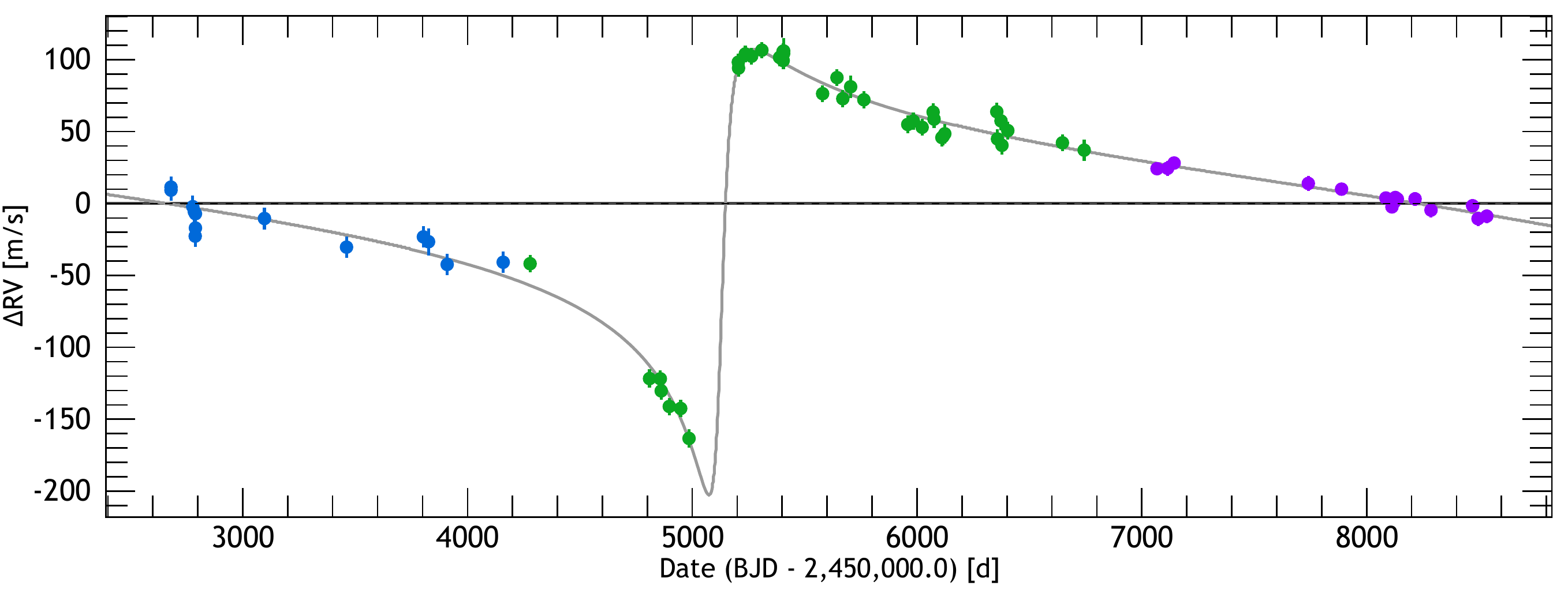}}
  \resizebox{\hsize}{!}{\includegraphics{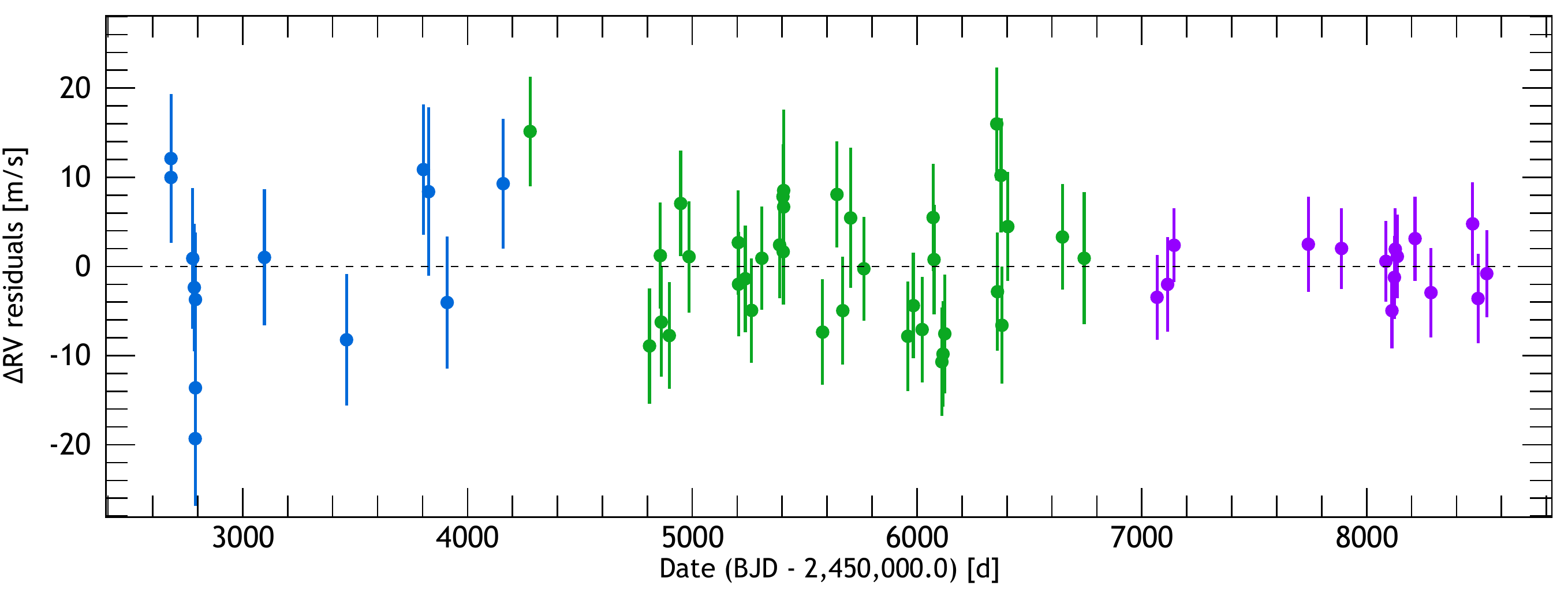}}
  \resizebox{\hsize}{!}{\includegraphics{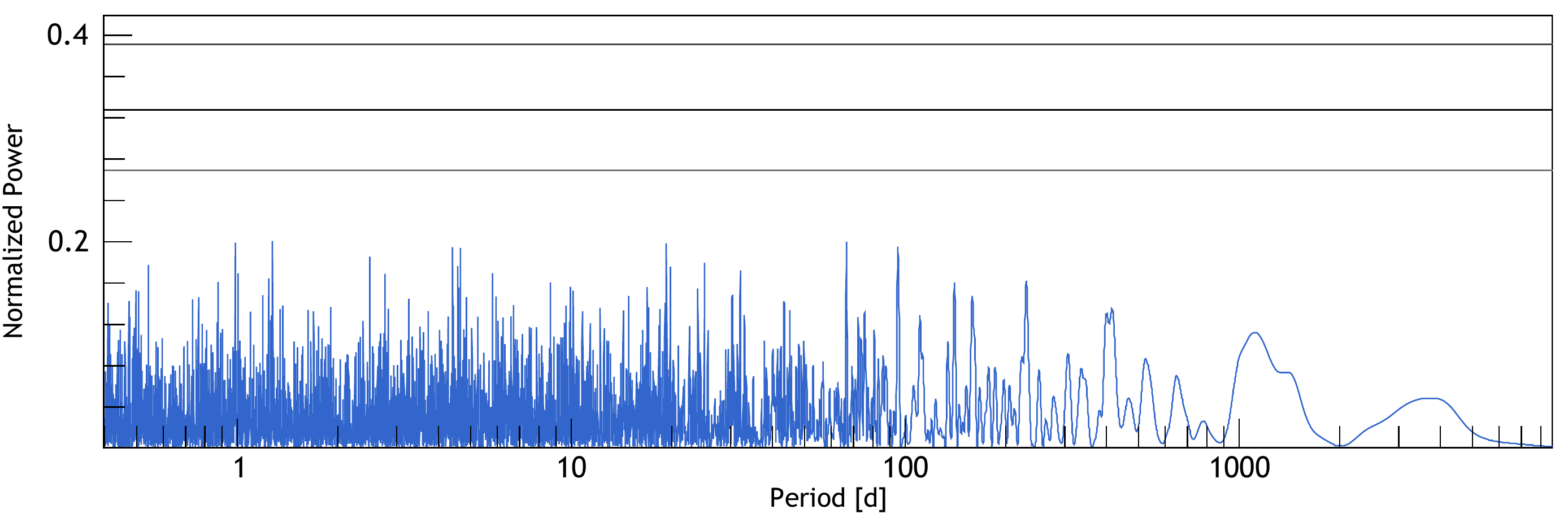}}
      \caption{\textbf{Top:} HD~98649 Radial velocity measurements as a function of Julian Date obtained with CORALIE-98 (blue), CORALIE-07 (green) and CORALIE-14 (purple). The best single-planet Keplerian model is represented as a black curve. \textbf{Middle:} The RV residuals of HD~98649.
      \textbf{Bottom:} The periodogram of the residuals for HD~98649 showing no significant signals remaining. The three black lines represent the $10 \%$, $1 \%$ and $0.1 \%$ false alarm probability in ascending order.}
      \label{fig:5}
\end{figure}

We report updated parameters for a known exoplanet detected by \cite{2013A&A...551A..90M}. HD~98649 is a G3/G5V star at $42.19 \pm 0.09$ pc from the Sun. The star properties are summarised in Table~\ref{table:4}.

HD~98649 has been observed with CORALIE at La Silla Observatory since February 2003. Fourteen measurements were taken with CORALIE-98, 42 additional radial velocity measurements were obtained with CORALIE-07, followed by 12 additional radial velocity measurements were obtained with CORALIE-14.

It is an extremely eccentric planet with an eccentricity of $0.86^{+0.04}_{-0.02}$. The orbital parameters agree well with \cite{2013A&A...551A..90M} who reported a period of $P = 13.56^{+1.66}_{-1.27}$ years and a mass of $M \sin i = 6.8 \pm 0.5\ M_{\text{Jup}}$.

The orbital solutions for HD~98649 are summarised in Table~\ref{table:4}. Figure~\ref{fig:5} shows the CORALIE radial velocities and the corresponding best-fit Keplerian model along with the radial velocity residuals and a periodogram of the residuals. The results from the fully probed parameter space from the MCMC are shown in the appendix.

\subsection{HD~50499 (HIP~32970)}

\begin{figure}
  \resizebox{\hsize}{!}{\includegraphics{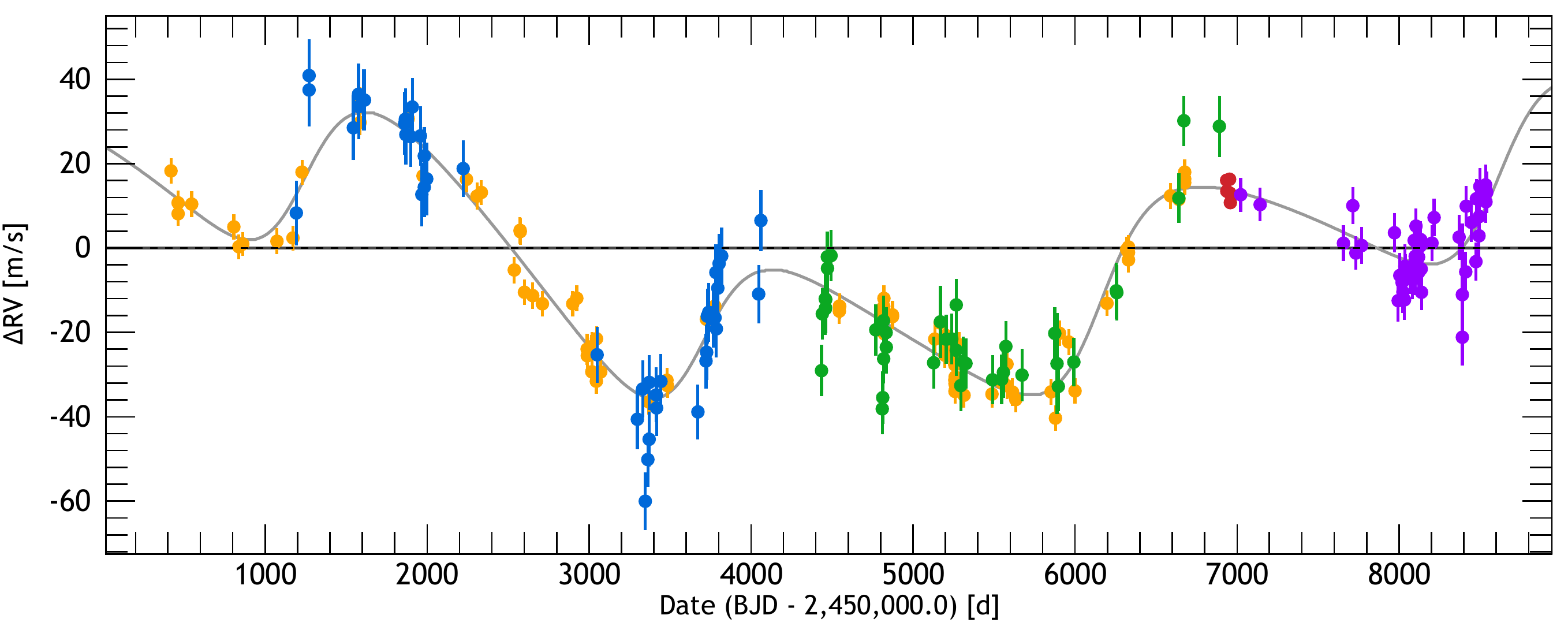}}
  \resizebox{\hsize}{!}{\includegraphics{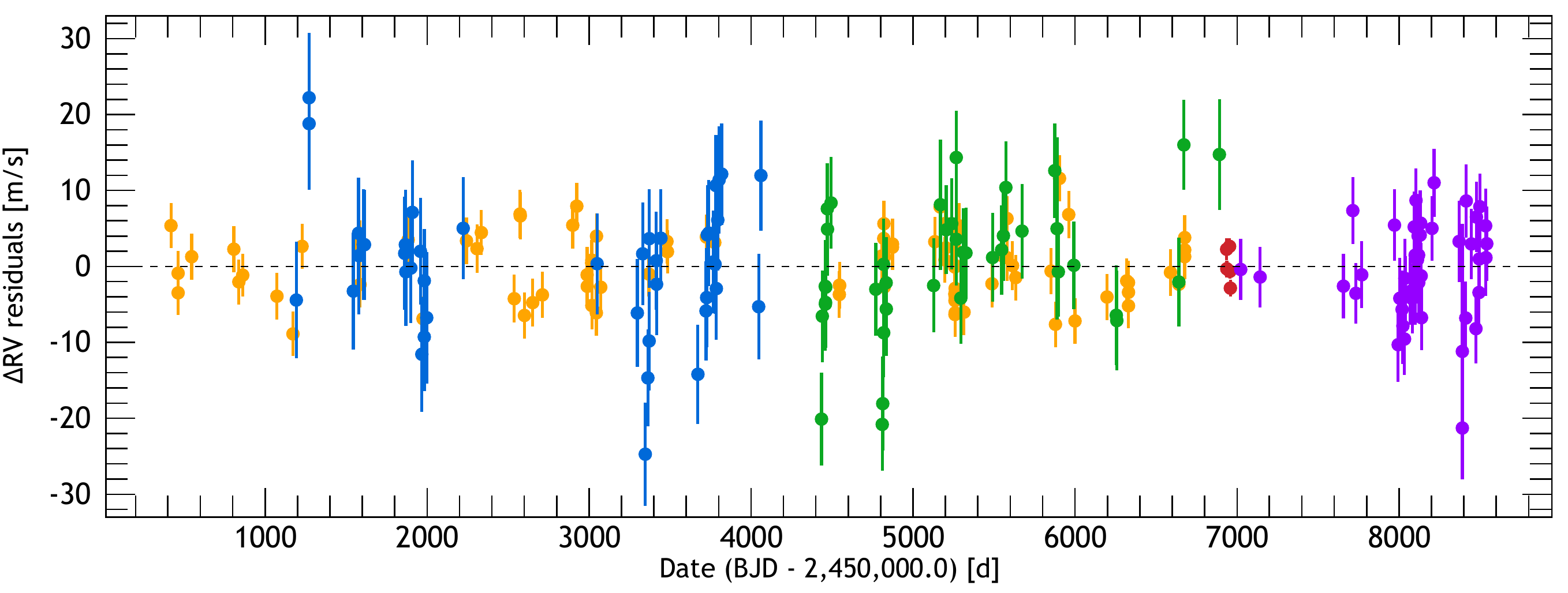}}
    \resizebox{\hsize}{!}{\includegraphics{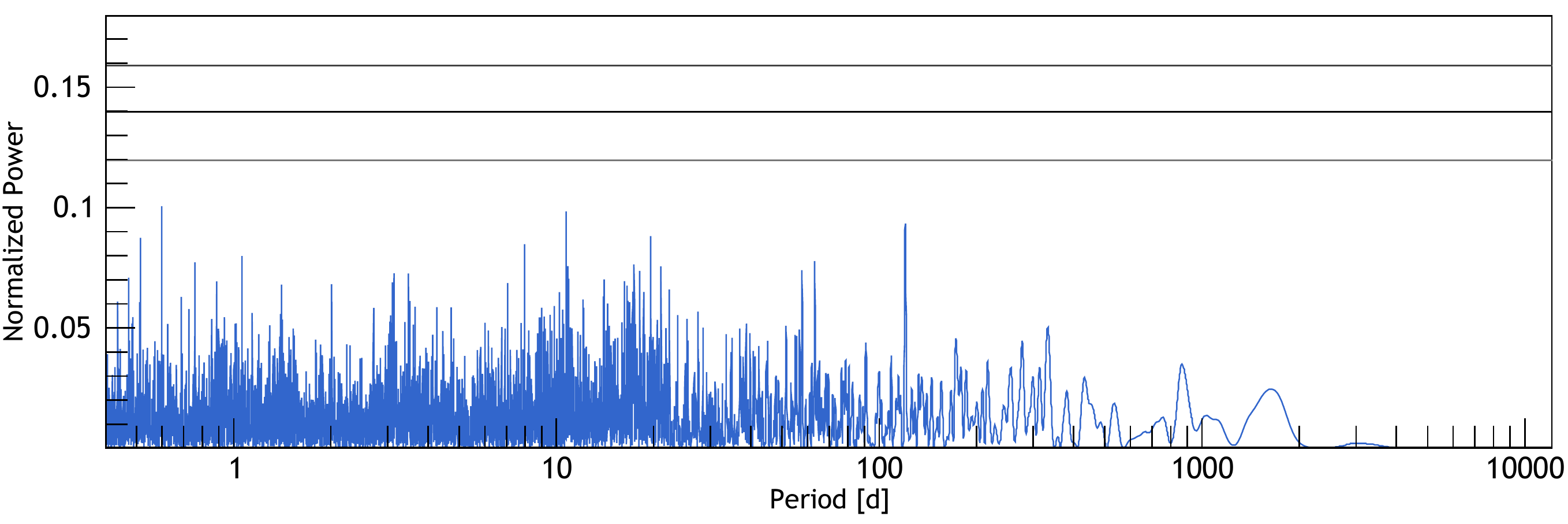}}
  \caption{\textbf{Top:} HD~50499 radial velocity curves. Blue: CORALIE-98 data; green: CORALIE-07; purple: CORALIE-14 data; red: HARPS data; orange: HIRES data \citep{2017AJ....153..208B}. The Keplerian models are represented as the black curves. \textbf{Middle:} The RV residuals of HD~50499. \textbf{Bottom:} Periodogram of the residuals of HD~50499 after the two planetary signals have been removed, indicating that there are no more significant signals remaining in the data. The three black lines represent the $10 \%$, $1 \%$ and $0.1 \%$ false alarm probability in ascending order.}
  \label{fig:6}
\end{figure}

\begin{figure}
  \resizebox{\hsize}{!}{\includegraphics{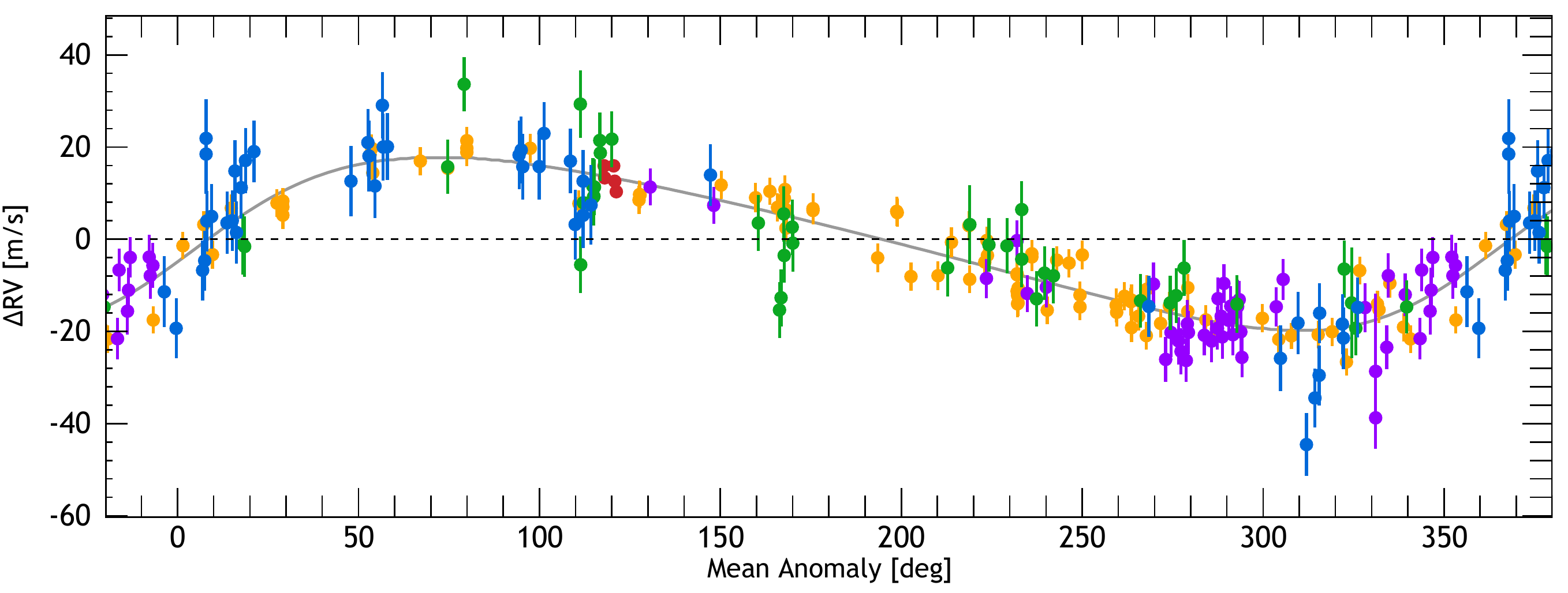}}
    \resizebox{\hsize}{!}{\includegraphics{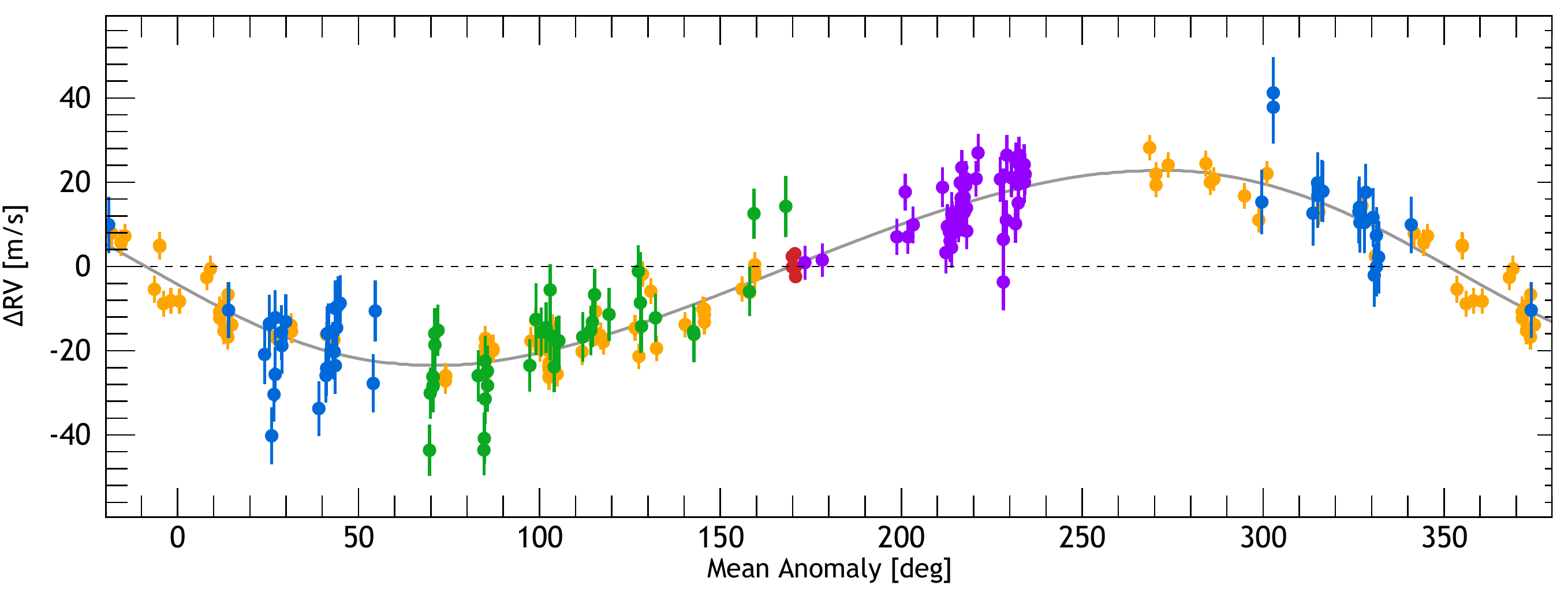}}
  \caption{Phase-folded radial velocity curves for HD~50499. Blue: CORALIE-98 data; green: CORALIE-07; purple: CORALIE-14 data; red: HARPS data; orange: HIRES data. The Keplerian models are represented as the black curves. \textbf{Top:} Phase-folded curve for HD~50499b. \textbf{Bottom:} Phase-folded curve for HD~50499c.}
  \label{fig:7}
\end{figure}

\begin{table*}
\caption{Best-fitted solution for the substellar companions orbiting HD~50499, HD~92788 and HD~98649. For each parameter, the mode of the posterior is considered, with error bars computed from the MCMC chains with 10,000,000 iterations using a 68.27\% confidence interval.}     % title of Table
\label{table:4}      % is used to refer this table in the text
\centering           % used for centering table
\begin{tabular}{c c c c c c c}        % centered columns (4 columns)
\hline\hline                    % inserts double horizontal lines
Parameters & Units & HD~50499b & HD~50499c & HD~92788b & HD~92788c & HD~98649b \\    % table heading
\hline                          % inserts single horizontal line
    $P$ & $\text{[years]}$ & $6.80 \pm 0.05$ & $23.6^{+7.18}_{-1.11}$ & $0.892 \pm 0.0001$ & $31.79^{+13.84}_{-2.48}$ & $16.49^{+1.13}_{-0.70}$ \\
    $K$ & $\text{[ms}^{-1}\text{]}$ & $18.94^{+0.82}_{-0.86}$ & $24.23^{+3.79}_{-0.95}$ & $108.24^{+0.89}_{-0.84}$ & $33.29^{+2.33}_{-1.94}$ & $140.1^{+33.1}_{-6.1}$ \\
    $e$ & & $0.27^{+0.04}_{-0.03}$ & $0.00^{+0.14}_{-0.02}$ & $0.35^{+0.004}_{-0.005}$ & $0.46^{+0.12}_{-0.03}$ & $0.86^{+0.04}_{-0.02}$ \\
    $\omega$ & $[\text{deg}]$ & $259.32^{+7.89}_{-10.19}$ & $[-115,+161]$ & $-82.17^{+1.01}_{-1.16}$ & $-25.71^{+6.63}_{-8.92}$ & $252.61^{+1.97}_{-7.03}$ \\
    $T_p$ & $[\text{JD}]$ & $6172.9^{+50.4}_{-67.5}$ & $11832^{+3731}_{-2885}$ & $5647.14^{+0.73}_{-0.73}$ & $6858^{+133}_{-202}$ & $5121.7^{+16.8}_{-28.1}$ \\
    \hline
    $M. \sin i$ & $M_{\text{Jup}}$ & $1.45 \pm 0.08$ & $2.93^{+0.73}_{-0.18}$ & $3.76^{+0.16}_{-0.15}$ & $3.67^{+0.30}_{-0.25}$ & $6.79^{+0.53}_{-0.31}$ \\
    $a$ & $[\text{AU}]$ & $3.93 \pm 0.07$ & $9.02^{+1.73}_{-0.33}$ & $0.97 \pm 0.02$ & $10.50^{+2.90}_{-0.55}$ & $6.57^{+0.31}_{-0.23}$ \\
    \hline
    $N_{\text{RV}}$ & & \multicolumn{2}{c}{214} & \multicolumn{2}{c}{214} & 68 \\
    $\Delta T$ & $[\text{years}]$ & \multicolumn{2}{c}{19.9} & \multicolumn{2}{c}{18.8} & 15.8 \\
\hline                              %inserts single line
\end{tabular}
\tablefoot{
$\Delta T$ is the time interval between the first and last measurements. C98 stands for CORALIE-98, C07 for CORALIE-07 and C14 for CORALIE-14. $N_{\text{RV}}$ is the number of RV measurements. Because the $\omega$ for HD~50499c is not very well constrained, we just provide the 68.27$\%$ confidence interval here. The fully probed parameters from the MCMC can be found in the appendix. $T_P$ is shown in BJD-2,450,000.}
\end{table*}

HD~50499 is a G1V star at $46.34 \pm 0.06$ pc from the Sun. The star properties are summarised in Table~\ref{table:1}.

We report updated orbital parameters for a known exoplanet around HD~50499b previously detected by \cite{2005ApJ...632..638V}. We also report the discovery of a new exoplanet in this system.

HD~50499 has been observed with CORALIE at La Silla Observatory since January 1999. Forty-four measurements were taken with CORALIE-98, 39 additional radial velocity measurements with CORALIE-07, followed by 40 additional radial velocity measurements with CORALIE-14. There are also 5 measurements taken with HARPS, as well as an additional 86 radial velocity points publicly available from HIRES.

The outer signal has been previously noted by \cite{2005ApJ...632..638V}. It has also been reported by \cite{2017AJ....153..208B} who noted that the outer trend is parabolic with additional data points from HIRES. In addition, \cite{2018arXiv180408329B} fits the radial velocity data for a single Keplerian orbit plus a quadratic term obtaining a best-fit curve where lower limits for the outer companion are derived. Here we provide additional data points from CORALIE, which allows us to further constrain this outer planet.

We derive the orbital solution for HD~50499 using two Keplerians as seen in Fig.~\ref{fig:6}. The orbital solutions for HD~50499 are summarised in Table~\ref{table:4}. Figure~\ref{fig:6} shows the CORALIE, HARPS and HIRES radial velocities and the corresponding best-fit Keplerian models along with the periodogram of the residuals. Figure~\ref{fig:7} shows the phase-folded radial velocity diagrams for HD~50499b and HD~50499c. The results from the fully probed parameter space from the MCMC are shown in the appendix.

These parameters agree well with \cite{2005ApJ...632..638V} who reports that HD~50499b has a period of $6.80 \pm 0.30$ years and a mass of $1.71 \pm 0.2\ M_{\mathrm{Jup}}$.

Because not all of the outer planet period is covered, the uncertainties on the period of the second Keplerian still remain relatively large with an unconstrained $\omega$.

The parameters we obtain for HD~50499c agree with \cite{2018arXiv180408329B} who report that the quadratic trend corresponds to a planet with an orbital period greater $P \geq 22.61$ years and a minimum mass of $M \sin i \geq 0.942 M_{\text{Jup}}$.

\subsection{HD~92788 (HIP~52409)}

\begin{figure}
  \resizebox{\hsize}{!}{\includegraphics{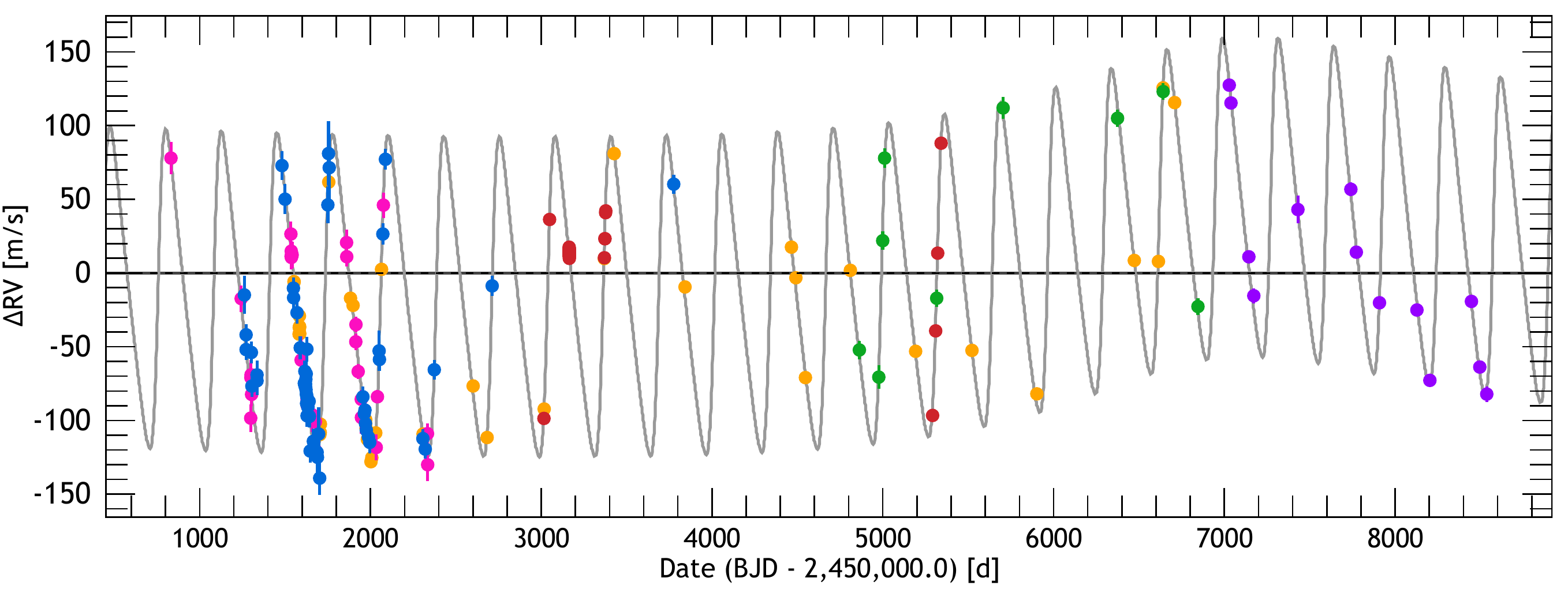}}
  \resizebox{\hsize}{!}{\includegraphics{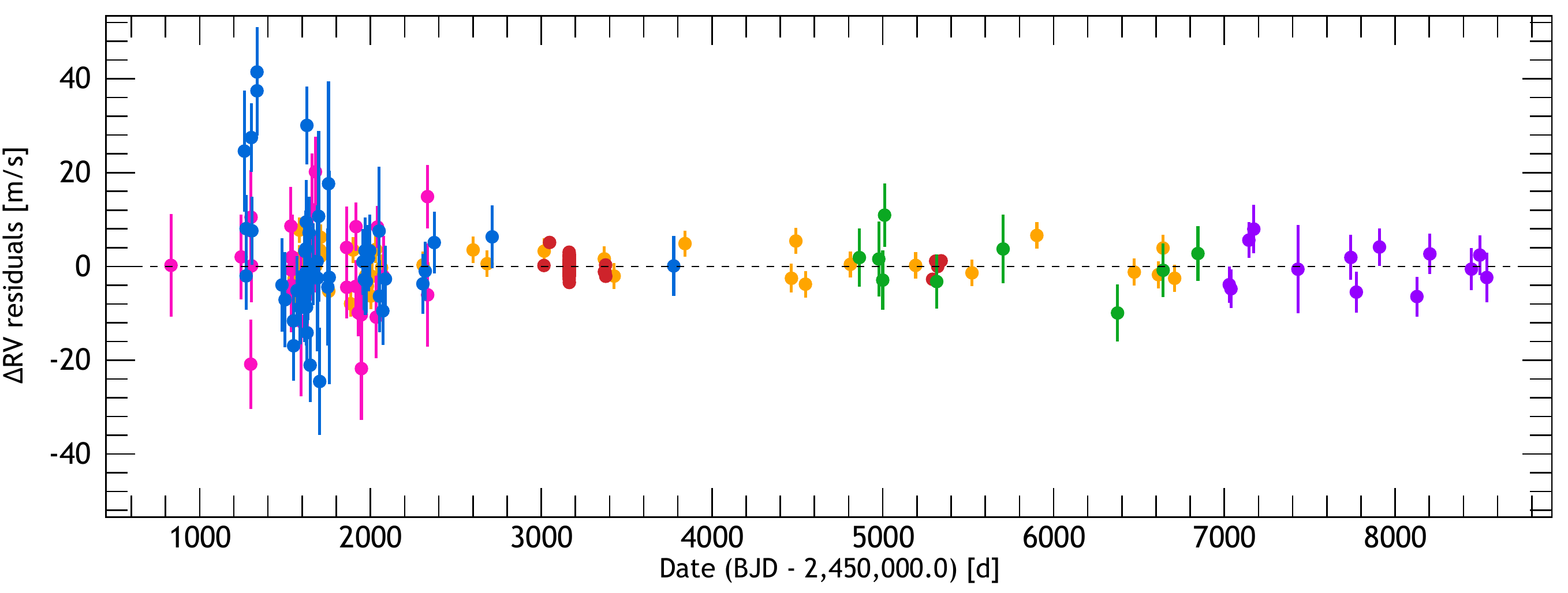}}
    \resizebox{\hsize}{!}{\includegraphics{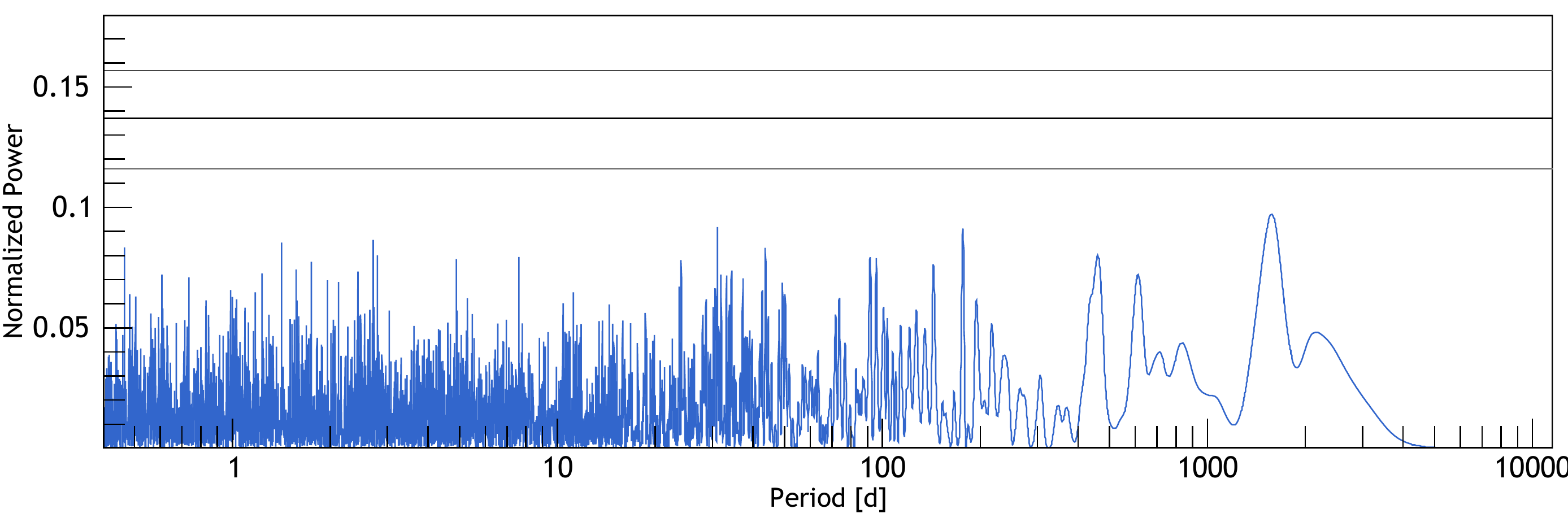}}
  \caption{\textbf{Top:} HD~92788 radial velocity measurements as a function of Julian Date obtained with CORALIE-98 (blue), CORALIE-07 (green), CORALIE-14 (purple), HARPS (red), HIRES (orange) \citep{2017AJ....153..208B} and HAMILTON (pink) \citep{2006ApJ...646..505B}. The best single-planet Keplerian model is represented as a black curve. \textbf{Middle:} The RV residuals of HD~92788. \textbf{Bottom:} The periodogram of the residuals for HD~92788 after the signal has been removed showing no significant signals. The three black lines represent the $10 \%$, $1 \%$ and $0.1 \%$ false alarm probability in ascending order.
  }
  \label{fig:8}
\end{figure}

\begin{figure}
    \resizebox{\hsize}{!}{\includegraphics{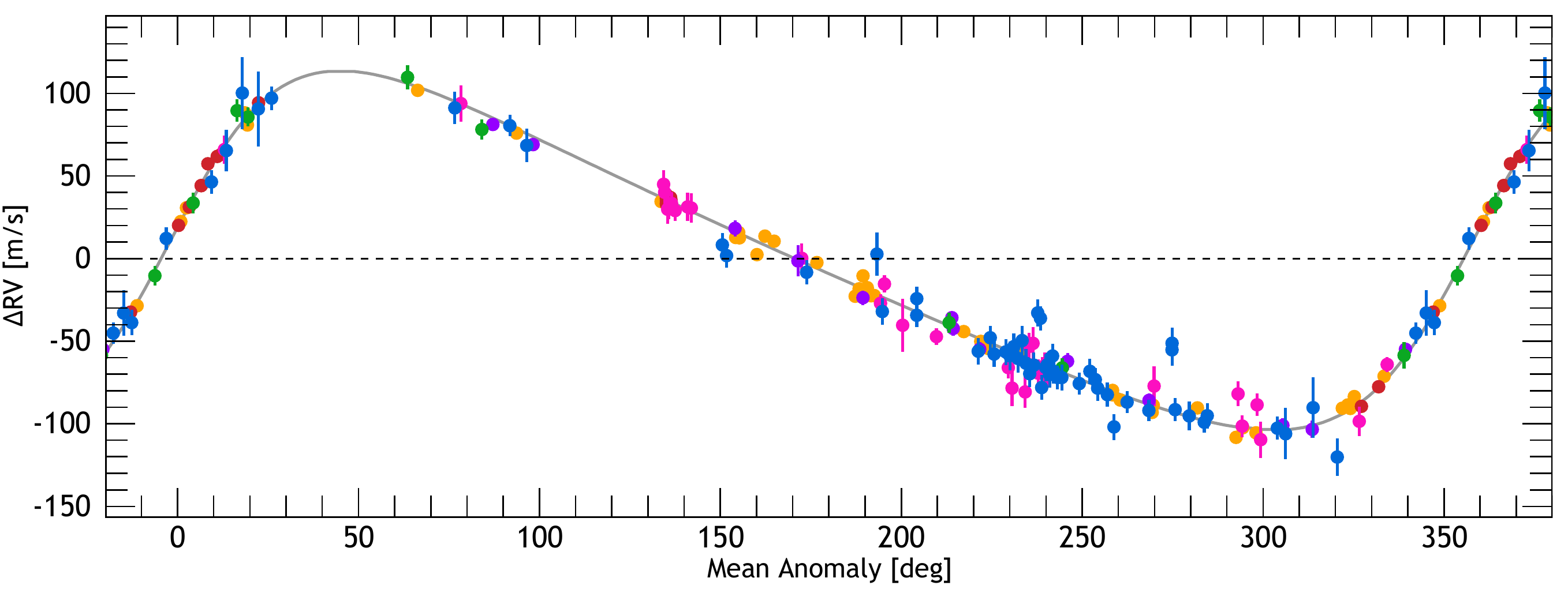}}
  \resizebox{\hsize}{!}{\includegraphics{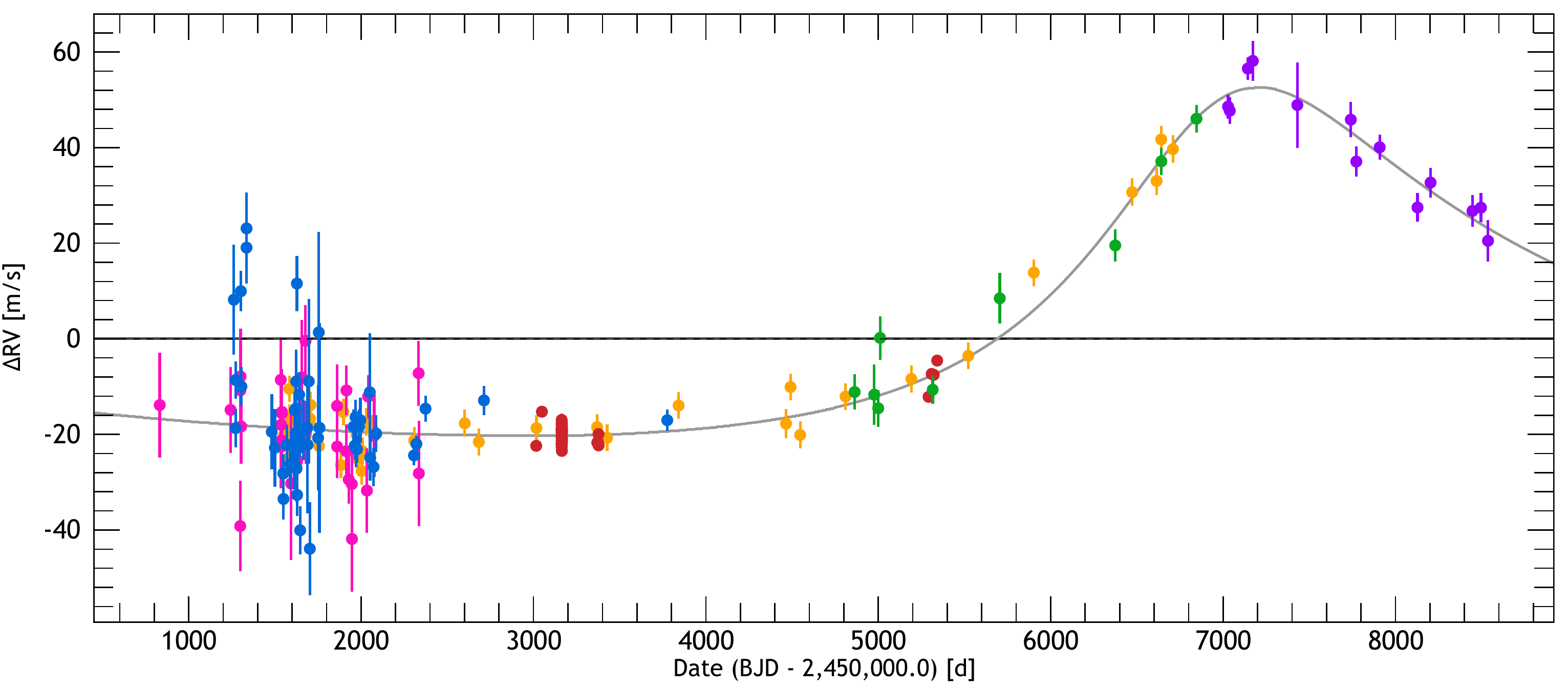}}
  \caption{Blue: CORALIE-98 data; green: CORALIE-07 data; purple: CORALIE-14 data; red: HARPS data; orange: HIRES data; pink: HAMILTON data. The Keplerian models are represented as the black curves. \textbf{Top:} Phase-folded curve for HD~92788b. \textbf{Bottom:} Time series for HD~92788c.}
  \label{fig:9}
\end{figure}

We report updated orbital parameters for HD~92788b \citep{2001ApJ...551.1107F} and confirm the detection of HD~92788c \citep{2019MNRAS.484.5859W}. HD~92788 is a G6V star at $28.83 \pm 0.05$ pc from the Sun. The star properties are summarised in Table~\ref{table:4}.

HD~92788 has been observed with CORALIE at La Silla Observatory since March 1999, 59 radial velocity measurements were obtained with CORALIE-98, an additional 10 radial velocity measurements were obtained with CORALIE-07 and an additional 11 radial velocity measurements were obtained with CORALIE-14. There are also 61 measurements taken with HARPS, as well as 42 radial velocity points publicly available from HIRES and an additional 31 radial velocity points from HAMILTON.

The orbital solutions for HD~92788 are summarised in Table~\ref{table:4}. Figure~\ref{fig:8} shows the CORALIE, HARPS, HIRES and HAMILTON radial velocities and the corresponding best-fit Keplerian models along with a time series of the residuals and a periodogram of the residuals. Figure~\ref{fig:9} shows the phase-folded radial velocity diagram for HD~92788b and the time series for HD~92788c. The results from the fully probed parameter space from the MCMC are shown in the appendix.

The orbital parameters for HD~92788b agree well with \cite{2001ApJ...551.1107F} who reported a period of $P = 0.894 \pm 0.009$ years and a minimum mass of $M \sin i = 3.34 M_{\mathrm{Jup}}$ and also agree well with the recently reported parameters by \cite{2019MNRAS.484.5859W} who report HD~92788b to have an orbital period of $0.892 \pm 0.00008$ years and a minimum mass of $M \sin i = 3.78 \pm 0.18 M_{\mathrm{Jup}}$.

\cite{2013ApJS..208....2W} tested the system for a potential additional planet with a period of 162 days which they found by limiting the eccentricity of the Keplerian model of HD~92788b. This is because fitting Keplerians can be biased towards fitting an increased eccentricity, especially when the semi-amplitude is small or the system has not been sampled well \citep{2008ApJ...685..553S}. Although \cite{2013ApJS..208....2W} suggested that there is a possible additional planet in the system, when we fitted the Keplerian the signal was not significant above the noise (lower than a $10 \%$ false alarm probability) to claim an additional planet at this period, as seen in Fig.~\ref{fig:8} and therefore we do not detect this signal.

We do, however, detect the signal of another planet in the system, HD~92788c, as recently discovered by \cite{2019MNRAS.484.5859W}. We confirm this planet with orbital parameters that agree with \cite{2019MNRAS.484.5859W}, who report a period of $26.99 \pm 2.54$ years and a minimum mass of $M. \sin i$ of $3.64 \pm 0.69 M_{\mathrm{Jup}}$.

\section{Discussion and Conclusion} \label{sec:5}

We have reported in this paper the discovery of five new giant planets and brown dwarfs candidates discovered with the CORALIE spectrograph mounted on 1.2m Euler Swiss telescope at La Silla Observatory as well as updated orbital parameters for four previously detected planets. CORALIE time series combined with the published data sets independently confirms the existence of these four already published companions. In addition, we do not find any significant evidence for the exoplanet suggested by \cite{2013ApJS..208....2W}. The newly reported companions span a period range of 15.6 to 40.4 years and a mass domain of 2.93 to 26.77 $M_{\mathrm{Jup}}$.

\begin{figure}
  \resizebox{\hsize}{!}{\includegraphics{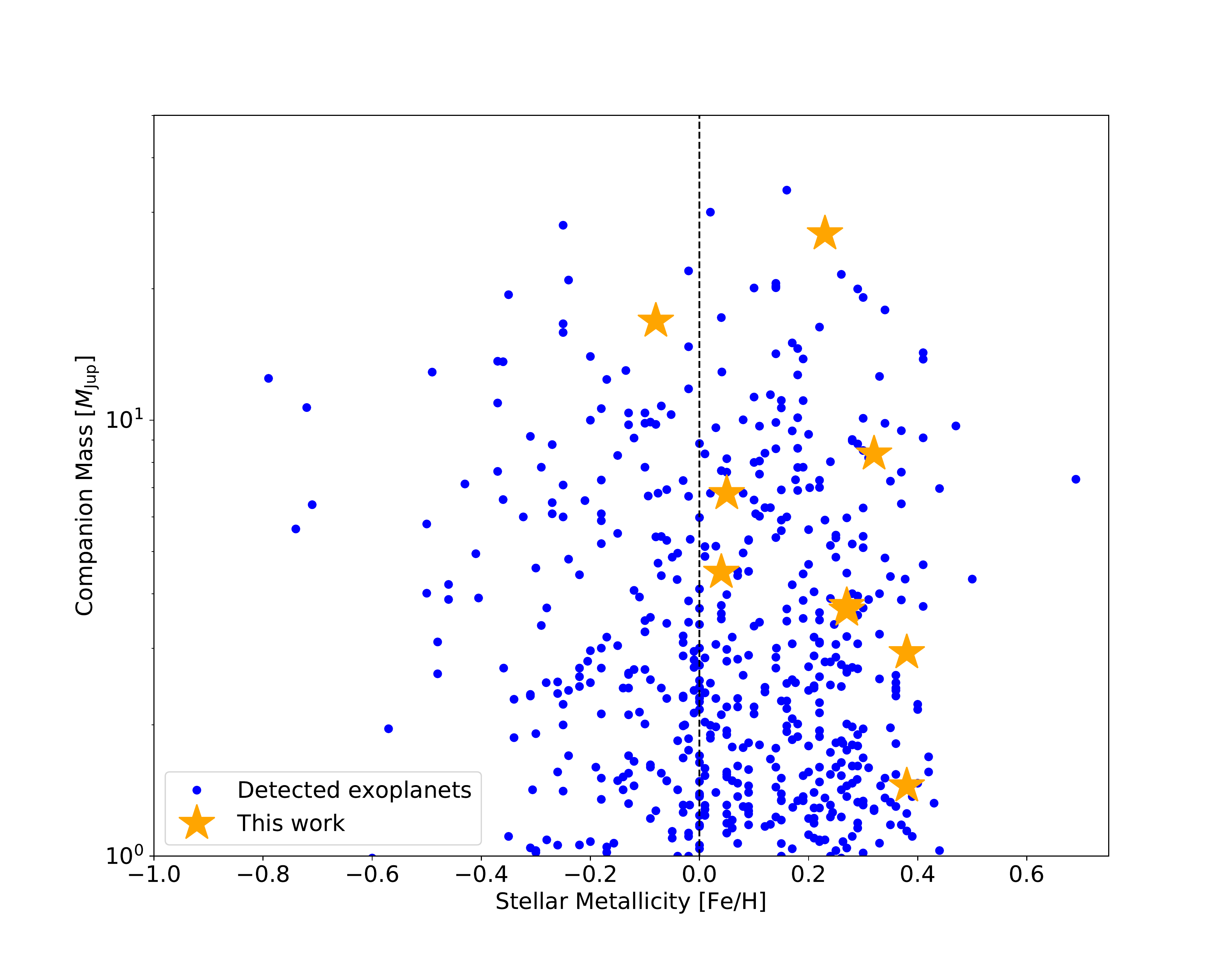}}
  \caption{Companion mass in the limits 1-50$M_{\mathrm{Jup}}$ as a function of the host star metallicity. The new companions presented in this paper are shown by the orange stars. Previously detected planets and brown dwarfs are shown in blue\protect\footnote. The black dashed line shows the metallicity of the Sun. Most of the stars with detected companions in this paper have a significant metallicity excess.}
  \label{fig:10}
\end{figure}

\footnotetext{Companions taken from the NASA Exoplanet Archive.}

Most of the parent stars in this paper have a metallicity excess, as seen in Figure~\ref{fig:10}. Despite the small size of our sample, our results seem to agree with previous observations that giant planets appear to occur significantly around stars that are more metal-rich, which has been noted before by \cite{2004A&A...415.1153S, 2005ApJ...622.1102F, 2011arXiv1109.2497M, 2012A&A...545A..55B}.

The focus of this paper has been on long period exoplanets, where all of the newly reported planets have periods over 15 years. This contributes to adding to the relatively small number of previously known planets with periods in this range, where according to the NASA Exoplanet Archive\footnote{The NASA Exoplanet Archive can be accessed at \url{https://exoplanetarchive.ipac.caltech.edu/index.html}.} there are only 26 known exoplanets with a period greater than 15 years.

\begin{figure}
  \resizebox{\hsize}{!}{\includegraphics{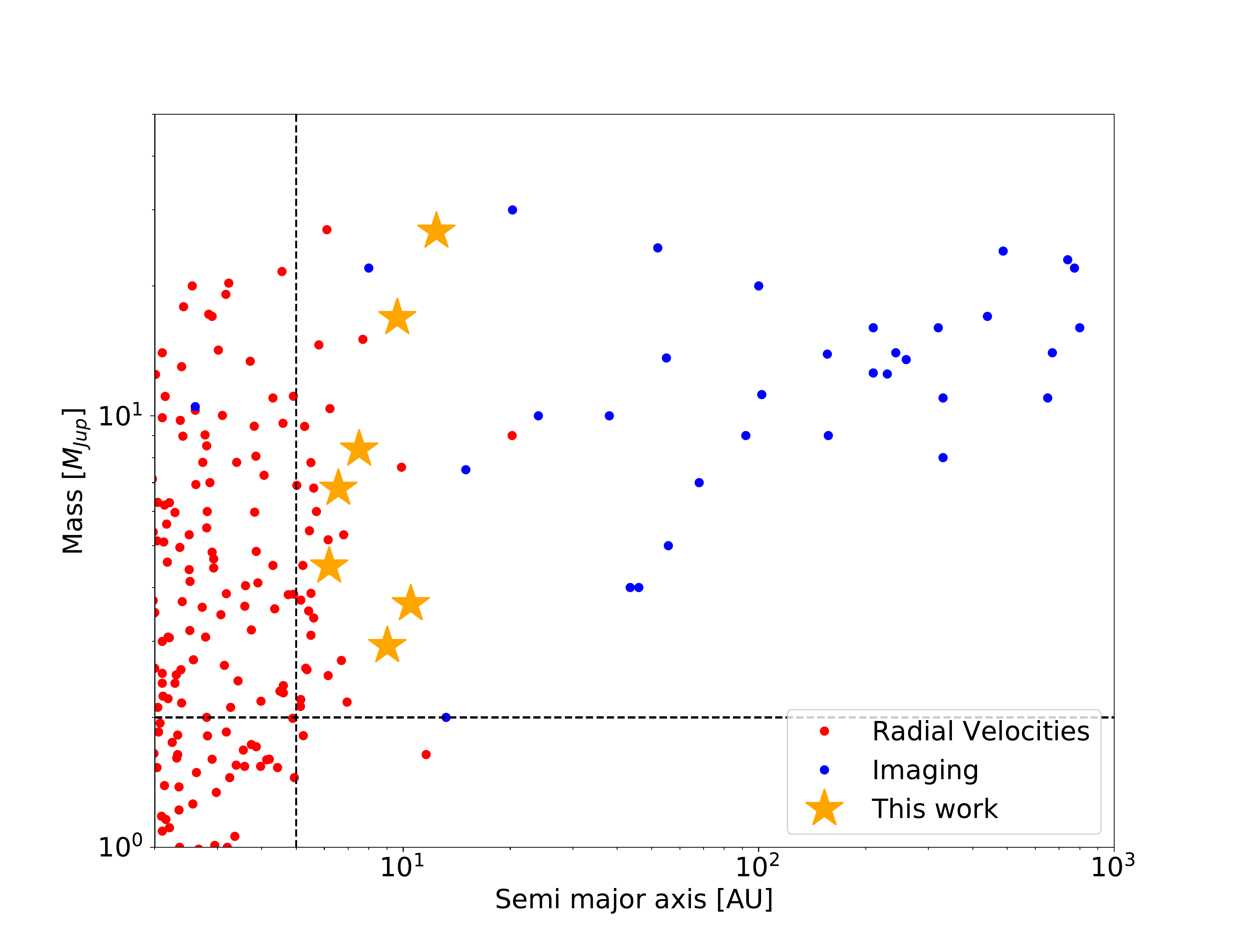}}
  \caption{Mass of detected exoplanets and brown dwarfs as a function of separation. The new companions presented in this paper are shown by the orange stars. Previously detected imaged planets and brown dwarfs are shown in blue and the radial velocity detected planets are shown in red$^2$. The black dashed lines show the limits for companions that are in a potential detectable parameter space with imaging, i.e. giant planets more massive than $2 M_{\text{Jup}}$ and planets at a separation larger than 5 AU.}
  \label{fig:11}
\end{figure}

As seen in Fig.~\ref{fig:11}, the planets and brown dwarfs presented in this paper are bridging the gap between the radial velocity detected exoplanets and the directly imaged exoplanets. As we achieve deeper detection limits and smaller inner working angles in imaging with new instrumentation and telescopes, as well as span longer base lines with radial velocity techniques, this gap in separation will decrease.

Combining radial velocity and direct imaging data has previously been utilised, for example by \cite{2014ApJ...785...93K}; \cite{2016ApJ...831..177R}; \cite{2016ApJ...818..106R} and \cite{2012ApJ...761...39C}. And more recently, combining these two techniques has been used with two candidates from the CORALIE RV survey with the detection of HD~4747~B by \cite{2018arXiv180505645P} and with the discovery of an ultra-cool brown dwarf companion to HD~4113~A by \cite{2018A&A...614A..16C}.

With direct imaging detections from these candidates in this paper, we aim to follow the same method as \cite{2018A&A...614A..16C} and \citep{2018arXiv180505645P}, and perform an atmospheric retrieval analysis of carbon and oxygen abundances, as demonstrated by \cite{2017AJ....154...91L}. The CORALIE survey plays an important step in identifying promising targets for such observations, making CORALIE a unique instrument in being able to carry out such a long continuous survey at high precision.b Furthermore, the stars in the CORALIE sample are older than typically imaged directly-imaged targets, and so the substellar companions probed in this paper represent a new and complementary parameter space.

Trying to detect these long period companions through imaging has been a part of a 15 year effort using VLT/NACO with little success. Now with the capabilities of VLT/SPHERE with a contrast limit of $\sim10^{-3}$ -- $10^{-4}$ at a separation of 0.1 arcseconds \citep{2019arXiv190204080B}, we are able to start detecting these massive planets and brown dwarfs.

Some of these targets may be challenging for SPHERE, but where this is the case, they should be within the capabilities of ELT/METIS, where METIS should achieve a $10^{-5}$ contrast at 0.1 arcseconds \citep{2016SPIE.9909E..73C}. In addition, combining astrometry, that will be available from \emph{Gaia}, with radial velocity data will allows us to further constrain the range of possible masses of these massive companions.

\begin{acknowledgements}
This work has been carried out within the framework of the National Centre for Competence in Research PlanetS supported by the Swiss National Science Foundation. The authors acknowledge the financial support of the SNSF. \newline

This publications makes use of the The Data \& Analysis Center for Exoplanets (DACE), which is a facility based at the University of Geneva (CH) dedicated to extrasolar planets data visualisation, exchange and analysis. DACE is a platform of the Swiss National Centre of Competence in Research (NCCR) PlanetS, federating the Swiss expertise in Exoplanet research. The DACE platform is available at https://dace.unige.ch. \newline 

LAdS and JVS acknowledge the support from the European Research Council (ERC) under the European Union’s Horizon 2020 research and innovation programme (project {\sc Four Aces}, grant agreement No 724427). \newline

NCS was supported by FCT - Fundação para a Ciência e a Tecnologia through national funds and by FEDER through COMPETE2020 - Programa Operacional Competitividade e Internacionalização by these grants: UID/FIS/04434/2013 \& POCI-01-0145-FEDER-007672; PTDC/FIS-AST/28953/2017 \& POCI-01-0145-FEDER-028953 and PTDC/FIS-AST/32113/2017 \& POCI-01-0145-FEDER-032113. \newline

This work has made use of data from the European Space Agency (ESA) mission \emph{Gaia} (\url{http://www.cosmos.esa.int/gaia}), processed by the \emph{Gaia} Data Processing and Analysis Consortium (DPAC, \url{http://www.cosmos.esa.int/web/gaia/dpac/consortium}). Funding for the DPAC has been provided by national institutions, in particular the institutions participating in the \emph{Gaia} Multilateral Agreement.
\end{acknowledgements}

\bibliographystyle{aa}
\bibliography{main.bbl}

\begin{appendix}
\section{Direct access to the radial velocities and other data products}
The radial velocity measurements and additional data products discussed in this paper are available in electronic form on the DACE web platform for each individual target with each link: \\

\begin{itemize}
\item HD~13724 \\
https://dace.unige.ch/radialVelocities/?pattern=HD13724
\item HD~181234 \\
https://dace.unige.ch/radialVelocities/?pattern=HD181234
\item HD~25015 \\
https://dace.unige.ch/radialVelocities/?pattern=HD25015
\item HD~50499 \\
https://dace.unige.ch/radialVelocities/?pattern=HD50499
\item HD~92788 \\
https://dace.unige.ch/radialVelocities/?pattern=HD92788
\item HD~92987 \\
https://dace.unige.ch/radialVelocities/?pattern=HD92987
\item HD~98649 \\
https://dace.unige.ch/radialVelocities/?pattern=HD98649
\end{itemize}

\section{MCMC Tables}

We probed the model parameter space with the MCMC sampler described in \cite{2014MNRAS.441..983D} and \cite{2016A&A...585A.134D}. For each MCMC simulation, we performed 10,000,000 iterations with initial conditions drawn from the solution obtained using \emph{DACE}. The corresponding parameters and confidence intervals for each star are listed in Tables \ref{tab:HD181234_mcmc-allparams}, \ref{tab:HD92987_mcmc-allparams}, \ref{tab:HD13724_mcmc-allparams}, \ref{tab:HD25015_mcmc-allparams}, \ref{tab:HD98649_mcmc-allparams}, \ref{tab:HD50499_mcmc-allparams} and \ref{tab:HD92788_mcmc-allparams} for HD~181234, HD~92987, HD~13724, HD~25015, HD~98649, HD~50499 and HD~92788 respectively.

\begin{sidewaystable*}
\scriptsize
  \begin{center}
    \caption{Parameters probed by the MCMC used to fit the RV measurements of HD~181234.
      The maximum likelihood solution, median, mode,
      and standard-deviation of the posterior distribution
      for each parameter are shown, as well as the 68.27\%,
      95.45\%, and 99.73\% confidence intervals.
      The prior for each parameter can be of type: $\mathcal{U}$:~uniform,
      $\mathcal{N}$:~normal, or $\mathcal{TN}$:~truncated normal.
      Reference epoch: 2455500.0~BJD.}
    \label{tab:HD181234_mcmc-allparams}
    \begin{tabular}{cccccccccccc}
      \hline
      Parameter & Units & Max(Likelihood) & Mode & Mean & Std & Median & 68.27\% & 95.45\% & 99.73\% & Prior\\
      \hline
      $\log$(Likelihood) &  & -359.29 & -365.26 & -366.31 & 2.83 & -365.95 & [-369.09--363.55] & [-372.93--361.77] & [-377.96--360.46] & --\\
        \hline
        \multicolumn{11}{c}{\textbf{Star}}\\
        \hline
      $M_\mathrm{{S}}$ & [$M_\odot$] & 1.0495 & 1.0117 & 1.0101 & 0.0601 & 1.0105 & [0.9501-1.0704] & [0.8899-1.1296] & [0.8287-1.1895] & $\mathcal{{U}}$\\
      $\Pi_\mathrm{{S}}$ & [mas] & 21.093 & 20.916 & 20.915 & 0.209 & 20.915 & [20.708-21.123] & [20.498-21.336] & [20.280-21.538] & $\mathcal{{U}}$\\
      \hline
      \multicolumn{11}{c}{\textbf{Offset}}\\
      \hline
      $\gamma_{\mathrm{COR98(DRS-3.3)}}$ & [m/s] & -8.36 & -5.47 & -5.41 & 2.74 & -5.44 & [-8.14--2.69] & [-10.79-0.22] & [-13.61-3.38] & $\mathcal{N}(0,4)$\\
      $\gamma_{\mathrm{HIRES(Pub-2017)}}$ & [m/s] & 46701.91 & 46703.49 & 46703.39 & 2.42 & 46703.40 & [46701.00-46705.79] & [46698.55-46708.24] & [46695.72-46711.05] & $\mathcal{{U}}$\\
      $\gamma_{\mathrm{COR07(DRS-3.4)}}$ & [m/s] & -46664.21 & -46664.26 & -46664.23 & 1.99 & -46664.23 & [-46666.21--46662.23] & [-46668.23--46660.32] & [-46670.43--46658.23] & $\mathcal{{U}}$\\
      $\gamma_{\mathrm{COR14(DRS-3.8)}}$ & [m/s] & 15.82 & 14.32 & 14.40 & 3.31 & 14.38 & [11.08-17.72] & [7.84-21.08] & [4.93-24.60] & $\mathcal{N}(12,4)$\\
      \hline
      \multicolumn{11}{c}{\textbf{Noise}}\\
      \hline
      $\sigma_{\mathrm{COR98(DRS-3.3)}}$ & [m/s] & 5.52 & 7.12 & 8.04 & 2.82 & 7.70 & [5.40-10.67] & [3.39-14.71] & [1.18-20.48] & $\mathcal{{U}}$\\
      $\sigma_{\mathrm{COR07(DRS-3.4)}}$ & [m/s] & 8.13 & 8.15 & 8.82 & 2.14 & 8.57 & [6.77-10.84] & [5.34-13.84] & [4.10-17.95] & $\mathcal{{U}}$\\
      $\sigma_{\mathrm{HIRES(Pub-2017)}}$ & [m/s] & 2.966 & 3.265 & 3.555 & 0.790 & 3.453 & [2.803-4.300] & [2.287-5.447] & [1.845-6.988] & $\mathcal{{U}}$\\
      $\sigma_{\mathrm{COR14(DRS-3.8)}}$ & [m/s] & 2.846 & 3.378 & 3.413 & 0.815 & 3.394 & [2.631-4.205] & [1.810-5.122] & [0.703-6.113] & $\mathcal{{U}}$\\
          \hline
          \multicolumn{11}{c}{\textbf{HD~181234b}}\\
          \hline
      $P$ & [d] & 7515.1 & 7463.4 & 7465.4 & 80.7 & 7464.3 & [7385.6-7545.4] & [7307.0-7632.1] & [7221.7-7718.5] & $\mathcal{{U}}$\\
      $K$ & [m/s] & 126.31 & 126.75 & 126.81 & 1.71 & 126.77 & [125.11-128.52] & [123.49-130.32] & [122.00-132.26] & $\mathcal{{U}}$\\
    $e$ &  & 0.72845 & 0.73177 & 0.73216 & 0.00710 & 0.73211 & [0.72512-0.73925] & [0.71793-0.74649] & [0.71028-0.75311] & $\mathcal{{U}}$\\ 
      $\omega$ & [$^\circ$] & 94.75 & 93.27 & 93.23 & 1.78 & 93.24 & [91.46-95.01] & [89.62-96.75] & [87.88-98.58] & $\mathcal{{U}}$\\
      $T_\mathrm{Vmin}$ & [BJD] & 2460630.87 & 2457946.63 & 2457946.37 & 8.55 & 2457946.40 & [2457937.94-2457954.82] & [2457928.94-2457963.58] & [2457919.90-2457972.50] & $\mathcal{{U}}$\\         \hline
      $a_\mathrm{S}$ & [AU] & 0.05978 & 0.05922 & 0.05926 & 0.00110 & 0.05925 & [0.05816-0.06036] & [0.05711-0.06153] & [0.05605-0.06275] & --\\
      $a$ & [AU] & 7.650 & 7.517 & 7.517 & 0.159 & 7.519 & [7.359-7.676] & [7.192-7.828] & [7.019-7.976] & --\\
      $m$ & [$M_\oplus$] & 2737 & 2659 & 2656 & 111 & 2657 & [2544-2767] & [2430-2877] & [2318-2986] & --\\
      $m$ & [$M_\mathrm{{J}}$] & 8.613 & 8.366 & 8.358 & 0.351 & 8.361 & [8.006-8.707] & [7.648-9.053] & [7.294-9.395] & --\\
      $m$ & [$M_\odot$] & 0.008222 & 0.007986 & 0.007978 & 0.000335 & 0.007980 & [0.007641-0.008311] & [0.007300-0.008641] & [0.006962-0.008968] & --\\
      $T_\mathrm{C}$ & [BJD] & 2460344.43 & 2457661.70 & 2457661.72 & 2.36 & 2457661.72 & [2457659.38-2457664.06] & [2457656.97-2457666.45] & [2457654.32-2457669.15] & --\\
      $T_\mathrm{P}$ & [BJD] & 2460355.10 & 2457668.66 & 2457668.91 & 5.26 & 2457668.85 & [2457663.69-2457674.13] & [2457658.52-2457679.58] & [2457653.25-2457685.35] & --\\
      \hline
    \end{tabular}
  \end{center}
\end{sidewaystable*}

\begin{sidewaystable*}
\scriptsize
  \begin{center}
    \caption{Parameters probed by the MCMC used to fit the RV measurements of HD~92987.
      The maximum likelihood solution, median, mode,
      and standard-deviation of the posterior distribution
      for each parameter are shown, as well as the 68.27\%,
      95.45\%, and 99.73\% confidence intervals.
      The prior for each parameter can be of type: $\mathcal{U}$:~uniform,
      $\mathcal{N}$:~normal, or $\mathcal{TN}$:~truncated normal.
      Reference epoch: 2455500.0~BJD.}
    \label{tab:HD92987_mcmc-allparams}
    \begin{tabular}{cccccccccccc}
      \hline
      Parameter & Units & Max(Likelihood) & Mode & Mean & Std & Median & 68.27\% & 95.45\% & 99.73\% & Prior\\
      \hline
      $\log$(Likelihood) &  & -287.21 & -292.23 & -293.24 & 2.60 & -292.89 & [-295.76--290.72] & [-299.43--289.15] & [-303.96--288.07] & --\\
        \hline
        \multicolumn{11}{c}{\textbf{Star}}\\
        \hline
      $M_\mathrm{{S}}$ & [$M_\odot$] & 1.0877 & 1.0795 & 1.0807 & 0.0601 & 1.0805 & [1.0207-1.1409] & [0.9602-1.2010] & [0.9018-1.2590] & $\mathcal{{U}}$\\
      $\Pi_\mathrm{{S}}$ & [mas] & 22.860 & 22.943 & 22.943 & 0.228 & 22.944 & [22.715-23.170] & [22.487-23.399] & [22.279-23.626] & $\mathcal{{U}}$\\
      \hline
      \multicolumn{11}{c}{\textbf{Offset}}\\
      \hline
      $\gamma_{\mathrm{COR14(DRS-3.8)}}$ & [m/s] & 11.89 & 11.28 & 11.21 & 3.40 & 11.21 & [7.80-14.61] & [4.35-18.01] & [0.77-21.33] & $\mathcal{N}(12,4)$\\
      $\gamma_{\mathrm{COR98(DRS-3.3)}}$ & [m/s] & 10.38 & 5.65 & 5.73 & 2.82 & 5.73 & [2.94-8.52] & [0.06-11.39] & [-2.92-14.40] & $\mathcal{N}(0,4)$\\
      $\gamma_{\mathrm{COR07(DRS-3.4)}}$ & [m/s] & 4757.07 & 4759.04 & 4759.37 & 2.21 & 4759.25 & [4757.21-4761.52] & [4755.31-4764.14] & [4753.41-4767.46] & $\mathcal{{U}}$\\
      \hline
      \multicolumn{11}{c}{\textbf{Noise}}\\
      \hline
      $\sigma_{\mathrm{COR07(DRS-3.4)}}$ & [m/s] & 2.88 & 3.55 & 3.76 & 1.60 & 3.66 & [2.25-5.26] & [0.71-7.32] & [0.03-10.30] & $\mathcal{{U}}$\\
      $\sigma_{\mathrm{COR98(DRS-3.3)}}$ & [m/s] & 5.408 & 5.746 & 5.968 & 0.925 & 5.899 & [5.065-6.876] & [4.315-8.029] & [3.656-9.431] & $\mathcal{{U}}$\\
      $\sigma_{\mathrm{COR14(DRS-3.8)}}$ & [m/s] & 2.22 & 2.68 & 2.79 & 1.40 & 2.73 & [1.38-4.13] & [0.26-5.85] & [0.02-8.18] & $\mathcal{{U}}$\\
          \hline
          \multicolumn{11}{c}{\textbf{HD~92987b}}\\
          \hline
    $\log P$ & [d] & 4.0204 & 4.0150 & 4.0208 & 0.0175 & 4.0191 & [4.0038-4.0376] & [3.9909-4.0607] & [3.9805-4.0921] & $\mathcal{{U}}$\\ 
    $\log K$ & [m/s] & 2.17745 & 2.18372 & 2.18316 & 0.00709 & 2.18324 & [2.17601-2.19025] & [2.16885-2.19710] & [2.16133-2.20394] & $\mathcal{{U}}$\\  
    $\sqrt{e}\cos\omega$ &  & -0.4573 & -0.4435 & -0.4426 & 0.0260 & -0.4428 & [-0.4682--0.4169] & [-0.4939--0.3898] & [-0.5227--0.3615] & $\mathcal{{U}}$\\
    $\sqrt{e}\sin\omega$ &  & -0.1091 & -0.1232 & -0.1119 & 0.0580 & -0.1154 & [-0.1694--0.0541] & [-0.2173-0.0143] & [-0.2573-0.0857] & $\mathcal{{U}}$\\
      $T_\mathrm{Vmin}$ & [BJD] & 2457596.8 & 2457600.4 & 2457599.0 & 26.3 & 2457599.7 & [2457573.1-2457625.1] & [2457543.7-2457649.7] & [2457513.1-2457674.0] & $\mathcal{{U}}$\\
       \hline
      $a_\mathrm{S}$ & [AU] & 0.14139 & 0.14234 & 0.14371 & 0.00487 & 0.14321 & [0.13900-0.14843] & [0.13550-0.15492] & [0.13232-0.16417] & --\\
      $a$ & [AU] & 9.687 & 9.621 & 9.672 & 0.318 & 9.653 & [9.362-9.981] & [9.093-10.365] & [8.849-10.841] & --\\
      $e$ &  & 0.2210 & 0.2109 & 0.2125 & 0.0164 & 0.2118 & [0.1966-0.2280] & [0.1817-0.2482] & [0.1657-0.2764] & --\\
      $K$ & [m/s] & 150.47 & 152.67 & 152.48 & 2.49 & 152.49 & [149.97-154.97] & [147.52-157.43] & [144.99-159.93] & --\\
      $\omega$ & [$^\circ$] & 193.42 & 195.10 & 194.29 & 7.60 & 194.55 & [186.68-201.84] & [178.30-208.69] & [169.69-214.64] & --\\
      $m$ & [$M_\oplus$] & 5312 & 5363 & 5369 & 214 & 5368 & [5155-5583] & [4939-5796] & [4727-6002] & --\\
      $m$ & [$M_\mathrm{{J}}$] & 16.714 & 16.876 & 16.893 & 0.673 & 16.891 & [16.222-17.566] & [15.541-18.238] & [14.874-18.887] & --\\
      $m$ & [$M_\odot$] & 0.015954 & 0.016108 & 0.016124 & 0.000643 & 0.016123 & [0.015484-0.016767] & [0.014834-0.017409] & [0.014197-0.018028] & --\\
      $P$ & [d] & 10482 & 10355 & 10499 & 428 & 10450 & [10089-10904] & [9794-11501] & [9560-12362] & --\\
      $T_\mathrm{C}$ & [BJD] & 2455569.4 & 2455533.3 & 2455534.2 & 40.4 & 2455533.6 & [2455493.9-2455574.5] & [2455455.4-2455617.2] & [2455416.6-2455662.2] & --\\
      $T_\mathrm{P}$ & [BJD] & 2457841 & 2457889 & 2457861 & 155 & 2457870 & [2457708-2458015] & [2457530-2458147] & [2457331-2458264] & --\\
      \hline
    \end{tabular}
  \end{center}

\end{sidewaystable*}

\begin{sidewaystable*}
\scriptsize
  \begin{center}
    \caption{Parameters probed by the MCMC used to fit the RV measurements of HD~13724.
      The maximum likelihood solution, median, mode,
      and standard-deviation of the posterior distribution
      for each parameter are shown, as well as the 68.27\%,
      95.45\%, and 99.73\% confidence intervals.
      The prior for each parameter can be of type: $\mathcal{U}$:~uniform,
      $\mathcal{N}$:~normal, or $\mathcal{TN}$:~truncated normal.
      Reference epoch: 2455500.0~BJD.}
    \label{tab:HD13724_mcmc-allparams}
    \begin{tabular}{cccccccccccc}
      \hline
      Parameter & Units & Max(Likelihood) & Mode & Mean & Std & Median & 68.27\% & 95.45\% & 99.73\% & Prior\\
      \hline
      $\log$(Likelihood) &  & -592.80 & -599.41 & -600.20 & 2.60 & -599.89 & [-602.74--597.67] & [-606.25--595.94] & [-610.74--594.59] & --\\
        \hline
        \multicolumn{11}{c}{\textbf{Star}}\\
        \hline
      $M_\mathrm{{S}}$ & [$M_\odot$] & 1.1744 & 1.1406 & 1.1397 & 0.0599 & 1.1396 & [1.0798-1.1994] & [1.0202-1.2596] & [0.9580-1.3200] & $\mathcal{{U}}$\\
      $\Pi_\mathrm{{S}}$ & [mas] & 23.306 & 22.978 & 22.979 & 0.230 & 22.979 & [22.749-23.208] & [22.520-23.441] & [22.302-23.663] & $\mathcal{{U}}$\\
      \hline
      \multicolumn{11}{c}{\textbf{Offset}}\\
      \hline
      $\gamma_{\mathrm{HARPS03(DRS-3.5)}}$ & [m/s] & 44.00 & 41.66 & 41.57 & 3.59 & 41.59 & [37.95-45.16] & [34.40-48.71] & [30.79-52.28] & $\mathcal{{U}}$\\
      $\gamma_{\mathrm{COR98(DRS-3.3)}}$ & [m/s] & 5.14 & 1.66 & 1.70 & 3.30 & 1.71 & [-1.58-4.99] & [-4.97-8.25] & [-8.27-11.42] & $\mathcal{N}(0,4)$\\
      $\gamma_{\mathrm{COR14(DRS-3.8)}}$ & [m/s] & 26.96 & 15.28 & 15.28 & 3.84 & 15.28 & [11.44-19.10] & [7.57-22.90] & [3.94-26.52] & $\mathcal{N}(12,4)$\\
      $\gamma_{\mathrm{COR07(DRS-3.4)}}$ & [m/s] & 20614.5 & 20597.2 & 20613.7 & 37.9 & 20608.9 & [20578.6-20647.3] & [20549.7-20716.3] & [20535.6-20735.2] & $\mathcal{{U}}$\\
      \hline
      \multicolumn{11}{c}{\textbf{Noise}}\\
      \hline
      $\sigma_{\mathrm{COR07(DRS-3.4)}}$ & [m/s] & 12.93 & 13.98 & 15.25 & 3.34 & 14.76 & [12.11-18.37] & [10.08-23.44] & [8.60-30.19] & $\mathcal{{U}}$\\
      $\sigma_{\mathrm{HARPS03(DRS-3.5)}}$ & [m/s] & 5.622 & 5.493 & 5.772 & 0.859 & 5.671 & [4.942-6.605] & [4.349-7.771] & [3.857-9.337] & $\mathcal{{U}}$\\
      $\sigma_{\mathrm{COR14(DRS-3.8)}}$ & [m/s] & 9.75 & 9.98 & 10.29 & 1.36 & 10.17 & [8.96-11.63] & [7.92-13.38] & [7.08-15.51] & $\mathcal{{U}}$\\
      $\sigma_{\mathrm{COR98(DRS-3.3)}}$ & [m/s] & 8.340 & 8.723 & 8.884 & 0.963 & 8.821 & [7.932-9.835] & [7.143-10.981] & [6.491-12.353] & $\mathcal{{U}}$\\
          \hline
          \multicolumn{11}{c}{\textbf{HD~13724b}}\\
          \hline
    $\log P$ & [d] & 4.2315 & 4.1727 & 4.2093 & 0.0979 & 4.1952 & [4.1196-4.2941] & [4.0453-4.4723] & [4.0079-4.5266] & $\mathcal{{U}}$\\
    $\log K$ & [m/s] & 2.3447 & 2.3310 & 2.3414 & 0.0330 & 2.3396 & [2.3098-2.3725] & [2.2780-2.4195] & [2.2580-2.4389] & $\mathcal{{U}}$\\
    $\sqrt{e}\cos\omega$ &  & -0.6095 & -0.5767 & -0.5881 & 0.0621 & -0.5862 & [-0.6446--0.5307] & [-0.7277--0.4578] & [-0.7525--0.4045] & $\mathcal{{U}}$\\
      $\sqrt{e}\sin\omega$ &  & -0.0557 & -0.0812 & -0.0824 & 0.0201 & -0.0818 & [-0.1020--0.0626] & [-0.1243--0.0439] & [-0.1538--0.0237] & $\mathcal{{U}}$\\
      $T_\mathrm{Vmin}$ & [BJD] & 2456047.7 & 2456038.8 & 2456038.6 & 28.8 & 2456039.2 & [2456009.9-2456067.5] & [2455979.1-2456093.7] & [2455943.9-2456120.6] & $\mathcal{{U}}$\\          \hline
      $a_\mathrm{S}$ & [AU] & 0.3212 & 0.2703 & 0.3157 & 0.0938 & 0.2954 & [0.2367-0.3865] & [0.1889-0.6070] & [0.1679-0.6801] & --\\
      $a$ & [AU] & 13.78 & 12.40 & 13.34 & 2.18 & 12.90 & [11.47-15.04] & [10.21-19.86] & [9.56-21.73] & --\\
      $e$ &  & 0.3745 & 0.3365 & 0.3569 & 0.0721 & 0.3507 & [0.2896-0.4215] & [0.2214-0.5347] & [0.1776-0.5692] & --\\
      $K$ & [m/s] & 221.1 & 214.3 & 220.1 & 16.9 & 218.6 & [204.1-235.8] & [189.7-262.7] & [181.1-274.7] & --\\
      $\omega$ & [$^\circ$] & 185.22 & 187.45 & 188.13 & 2.50 & 187.86 & [185.80-190.39] & [183.96-194.14] & [182.15-199.24] & --\\
      $m$ & [$M_\oplus$] & 9184 & 8507 & 8875 & 1126 & 8747 & [7810-9899] & [6972-11785] & [6404-12781] & --\\
      $m$ & [$M_\mathrm{{J}}$] & 28.90 & 26.77 & 27.93 & 3.54 & 27.52 & [24.58-31.15] & [21.94-37.08] & [20.15-40.22] & --\\
      $m$ & [$M_\odot$] & 0.02759 & 0.02555 & 0.02666 & 0.00338 & 0.02627 & [0.02346-0.02973] & [0.02094-0.03540] & [0.01923-0.03839] & --\\
      $P$ & [d] & 17040 & 14764 & 16637 & 4187 & 15673 & [13170-19684] & [11099-29669] & [10185-33620] & --\\
      $T_\mathrm{C}$ & [BJD] & 2453671 & 2453723 & 2453662 & 174 & 2453686 & [2453505-2453826] & [2453202-2453933] & [2452980-2454008] & --\\
      $T_\mathrm{P}$ & [BJD] & 2456152.0 & 2456189.1 & 2456199.0 & 66.5 & 2456195.2 & [2456135.4-2456261.8] & [2456076.2-2456345.7] & [2456008.8-2456458.5] & --\\
      \hline
    \end{tabular}
  \end{center}

\end{sidewaystable*}

\begin{sidewaystable*}
\scriptsize
  \begin{center}
    \caption{Parameters probed by the MCMC used to fit the RV measurements of HD~25015.
      The maximum likelihood solution, median, mode,
      and standard-deviation of the posterior distribution
      for each parameter are shown, as well as the 68.27\%,
      95.45\%, and 99.73\% confidence intervals.
      The prior for each parameter can be of type: $\mathcal{U}$:~uniform,
      $\mathcal{N}$:~normal, or $\mathcal{TN}$:~truncated normal.
      Reference epoch: 2455500.0~BJD.}
    \label{tab:HD25015_mcmc-allparams}
    \begin{tabular}{cccccccccccc}
      \hline
      Parameter & Units & Max(Likelihood) & Mode & Mean & Std & Median & 68.27\% & 95.45\% & 99.73\% & Prior\\
      \hline
      $\log$(Likelihood) &  & -430.28 & -435.20 & -436.10 & 2.51 & -435.77 & [-438.54--433.67] & [-442.07--432.17] & [-446.55--431.20] & --\\
        \hline
        \multicolumn{11}{c}{\textbf{Star}}\\
        \hline
      $M_\mathrm{{S}}$ & [$M_\odot$] & 0.8851 & 0.8598 & 0.8599 & 0.0502 & 0.8600 & [0.8097-0.9103] & [0.7591-0.9597] & [0.7098-1.0082] & $\mathcal{{U}}$\\
      $\Pi_\mathrm{{S}}$ & [mas] & 26.651 & 26.695 & 26.685 & 0.266 & 26.684 & [26.418-26.952] & [26.157-27.218] & [25.893-27.491] & $\mathcal{{U}}$\\
      \hline
      \multicolumn{11}{c}{\textbf{Offset}}\\
      \hline
      $\gamma_{\mathrm{COR07(DRS-3.4)}}$ & [m/s] & 28536.61 & 28537.76 & 28537.61 & 2.83 & 28537.63 & [28534.79-28540.43] & [28531.85-28543.20] & [28528.97-28545.98] & $\mathcal{{U}}$\\
      $\gamma_{\mathrm{COR14(DRS-3.8)}}$ & [m/s] & 15.62 & 12.81 & 12.77 & 3.80 & 12.76 & [8.96-16.57] & [5.21-20.38] & [1.55-24.36] & $\mathcal{N}(12,4)$\\
      $\gamma_{\mathrm{COR98(DRS-3.3)}}$ & [m/s] & 2.76 & -0.01 & -0.06 & 3.86 & -0.03 & [-3.93-3.79] & [-7.81-7.67] & [-11.51-11.69] & $\mathcal{N}(0,4)$\\
      \hline
      \multicolumn{11}{c}{\textbf{Activity cycle}}\\
      \hline
      $A_{H_\alpha\mathrm{, Gauss., 0.1 yr, low pass}}$ & [m/s] & 7.59 & 6.81 & 6.82 & 4.37 & 6.82 & [2.49-11.15] & [-1.95-15.58] & [-6.98-20.36] & $\mathcal{{U}}$\\
      \hline
      \multicolumn{11}{c}{\textbf{Noise}}\\
      \hline
      $\sigma_{\mathrm{COR07(DRS-3.4)}}$ & [m/s] & 10.26 & 12.07 & 10.77 & 4.79 & 11.17 & [5.57-15.48] & [0.95-19.66] & [0.05-24.44] & $\mathcal{{U}}$\\
      Low act. & [m/s] & 8.32 & 10.54 & 8.53 & 3.85 & 9.19 & [3.96-12.36] & [0.61-14.85] & [0.04-17.16] & $\mathcal{{U}}$\\
      $\sigma_{H_\alpha\mathrm{, Gauss., 0.1 yr, low pass}}$ & [m/s] & 14.60 & 14.31 & 13.98 & 6.73 & 14.01 & [6.90-20.61] & [1.20-27.89] & [0.06-35.04] & $\mathcal{{U}}$\\
      $\sigma_{\mathrm{COR98(DRS-3.3)}}$ & [m/s] & 16.12 & 16.47 & 16.35 & 5.92 & 16.40 & [10.75-21.88] & [3.47-28.45] & [0.21-36.46] & $\mathcal{{U}}$\\
      $\sigma_{\mathrm{COR14(DRS-3.8)}}$ & [m/s] & 5.53 & 8.28 & 6.48 & 3.87 & 6.41 & [2.03-10.71] & [0.28-13.87] & [0.02-17.05] & $\mathcal{{U}}$\\
          \hline
          \multicolumn{11}{c}{\textbf{HD~25015b}}\\
          \hline
      $\log P$ & [d] & 3.7879 & 3.7821 & 3.7937 & 0.0353 & 3.7888 & [3.7609-3.8259] & [3.7377-3.8807] & [3.7187-3.9585] & $\mathcal{{U}}$\\
    $\log K$ & [m/s] & 1.7817 & 1.7796 & 1.7788 & 0.0236 & 1.7791 & [1.7558-1.8016] & [1.7300-1.8254] & [1.7011-1.8552] & $\mathcal{{U}}$\\
    $\sqrt{e}\cos\omega$ &  & 0.190 & 0.133 & 0.128 & 0.100 & 0.128 & [0.027-0.228] & [-0.071-0.328] & [-0.171-0.432] & $\mathcal{{U}}$\\
      $\sqrt{e}\sin\omega$ &  & 0.5950 & 0.6088 & 0.6054 & 0.0631 & 0.6070 & [0.5449-0.6668] & [0.4712-0.7283] & [0.3832-0.7800] & $\mathcal{{U}}$\\ 
      $T_\mathrm{Vmax}$ & [BJD] & 2455258 & 2455209 & 2455210 & 122 & 2455209 & [2455090-2455329] & [2454964-2455458] & [2454827-2455591] & $\mathcal{{U}}$\\         \hline
      $a_\mathrm{S}$ & [AU] & 0.03141 & 0.03112 & 0.03144 & 0.00227 & 0.03132 & [0.02924-0.03361] & [0.02730-0.03639] & [0.02526-0.04051] & --\\
      $a$ & [AU] & 6.309 & 6.193 & 6.311 & 0.373 & 6.260 & [5.966-6.642] & [5.720-7.231] & [5.509-8.115] & --\\
      $e$ &  & 0.3901 & 0.3883 & 0.3969 & 0.0785 & 0.3944 & [0.3200-0.4745] & [0.2454-0.5622] & [0.1699-0.6497] & --\\
      $K$ & [m/s] & 60.49 & 60.11 & 60.18 & 3.29 & 60.13 & [56.99-63.33] & [53.71-66.90] & [50.25-71.65] & --\\
      $\omega$ & [$^\circ$] & 72.29 & 77.65 & 78.20 & 9.32 & 78.13 & [69.06-87.41] & [59.46-96.93] & [49.33-107.71] & --\\
      $m$ & [$M_\oplus$] & 1469.8 & 1424.3 & 1428.1 & 91.6 & 1427.0 & [1336.7-1520.0] & [1249.5-1614.0] & [1162.7-1713.7] & --\\
      $m$ & [$M_\mathrm{{J}}$] & 4.625 & 4.482 & 4.494 & 0.288 & 4.490 & [4.206-4.783] & [3.932-5.079] & [3.659-5.392] & --\\
      $m$ & [$M_\odot$] & 0.004415 & 0.004278 & 0.004289 & 0.000275 & 0.004286 & [0.004015-0.004565] & [0.003753-0.004848] & [0.003492-0.005147] & --\\
      $P$ & [d] & 6137 & 6021 & 6239 & 531 & 6149 & [5766-6697] & [5467-7598] & [5233-9090] & --\\
      $T_\mathrm{C}$ & [BJD] & 2455962.9 & 2455936.9 & 2455940.6 & 97.7 & 2455939.6 & [2455843.2-2456038.8] & [2455745.3-2456137.5] & [2455648.1-2456230.8] & --\\
      $T_\mathrm{P}$ & [BJD] & 2455840 & 2455852 & 2455861 & 149 & 2455858 & [2455715-2456009] & [2455568-2456166] & [2455413-2456323] & --\\
      \hline
    \end{tabular}
  \end{center}
\end{sidewaystable*}

\begin{sidewaystable*}
\scriptsize
  \begin{center}
    \caption{Parameters probed by the MCMC used to fit the RV measurements of HD~98649.
      The maximum likelihood solution, median, mode,
      and standard-deviation of the posterior distribution
      for each parameter are shown, as well as the 68.27\%,
      95.45\%, and 99.73\% confidence intervals.
      The prior for each parameter can be of type: $\mathcal{U}$:~uniform,
      $\mathcal{N}$:~normal, or $\mathcal{TN}$:~truncated normal.
      Reference epoch: 2455500.0~BJD.}
    \label{tab:HD98649_mcmc-allparams}
    \begin{tabular}{cccccccccccc}
      \hline
      Parameter & Units & Max(Likelihood) & Mode & Mean & Std & Median & 68.27\% & 95.45\% & 99.73\% & Prior\\
      \hline
      $\log$(Likelihood) &  & -201.70 & -207.30 & -208.21 & 2.60 & -207.89 & [-210.77--205.67] & [-214.31--204.03] & [-218.55--202.86] & --\\
        \hline
        \multicolumn{11}{c}{\textbf{Star}}\\
        \hline
      $M_\mathrm{{S}}$ & [$M_\odot$] & 0.9465 & 1.0296 & 1.0300 & 0.0599 & 1.0299 & [0.9701-1.0899] & [0.9105-1.1501] & [0.8513-1.2099] & $\mathcal{{U}}$\\
      $\Pi_\mathrm{{S}}$ & [mas] & 23.995 & 23.688 & 23.684 & 0.236 & 23.684 & [23.448-23.921] & [23.211-24.158] & [22.967-24.390] & $\mathcal{{U}}$\\
      \hline
      \multicolumn{11}{c}{\textbf{Offset}}\\
      \hline
      $\gamma_{\mathrm{COR98(DRS-3.3)}}$ & [m/s] & -7.05 & -1.94 & -1.87 & 3.41 & -1.87 & [-5.29-1.52] & [-8.66-5.01] & [-11.97-8.58] & $\mathcal{N}(0,4)$\\
      $\gamma_{\mathrm{COR14(DRS-3.8)}}$ & [m/s] & 20.61 & 16.33 & 16.18 & 2.86 & 16.22 & [13.33-19.03] & [10.38-21.83] & [7.29-24.66] & $\mathcal{N}(12,4)$\\
      $\gamma_{\mathrm{COR07(DRS-3.4)}}$ & [m/s] & 4281.50 & 4281.91 & 4282.18 & 2.27 & 4282.11 & [4279.94-4284.45] & [4277.87-4286.97] & [4275.82-4289.55] & $\mathcal{{U}}$\\
      \hline
      \multicolumn{11}{c}{\textbf{Noise}}\\
      \hline
      $\sigma_{\mathrm{COR98(DRS-3.3)}}$ & [m/s] & 8.66 & 8.89 & 9.87 & 2.82 & 9.46 & [7.25-12.47] & [5.51-16.73] & [4.17-23.36] & $\mathcal{{U}}$\\
      $\sigma_{\mathrm{COR14(DRS-3.8)}}$ & [m/s] & 0.62 & 0.46 & 1.75 & 1.29 & 1.51 & [0.47-3.00] & [0.07-4.90] & [0.00-7.73] & $\mathcal{{U}}$\\
      $\sigma_{\mathrm{COR07(DRS-3.4)}}$ & [m/s] & 5.63 & 6.49 & 6.73 & 1.15 & 6.64 & [5.61-7.85] & [4.73-9.32] & [3.95-11.07] & $\mathcal{{U}}$\\
          \hline
          \multicolumn{11}{c}{\textbf{HD~98649b}}\\
          \hline
    $\log P$ & [d] & 3.7508 & 3.7808 & 3.7848 & 0.0240 & 3.7834 & [3.7611-3.8086] & [3.7409-3.8373] & [3.7222-3.8692] & $\mathcal{{U}}$\\ 
    $\log K$ & [m/s] & 2.1918 & 2.1458 & 2.1817 & 0.0622 & 2.1650 & [2.1272-2.2386] & [2.1061-2.3537] & [2.0906-2.4329] & $\mathcal{{U}}$\\  
      $\sqrt{e}\cos\omega$ &  & -0.3301 & -0.2764 & -0.3155 & 0.0796 & -0.2974 & [-0.3900--0.2445] & [-0.5267--0.2028] & [-0.5957--0.1629] & $\mathcal{{U}}$\\
    $\sqrt{e}\sin\omega$ &  & -0.8689 & -0.8808 & -0.8715 & 0.0221 & -0.8769 & [-0.8885--0.8572] & [-0.8984--0.8016] & [-0.9072--0.7610] & $\mathcal{{U}}$\\
      $T_\mathrm{Vmax}$ & [BJD] & 2455271.6 & 2455264.7 & 2455262.2 & 14.1 & 2455263.5 & [2455250.4-2455274.8] & [2455227.6-2455286.4] & [2455190.8-2455299.0] & $\mathcal{{U}}$\\
        \hline
      $a_\mathrm{S}$ & [AU] & 0.04055 & 0.04129 & 0.04223 & 0.00280 & 0.04178 & [0.03969-0.04477] & [0.03789-0.04939] & [0.03626-0.05465] & --\\
      $a$ & [AU] & 6.097 & 6.570 & 6.609 & 0.277 & 6.594 & [6.336-6.881] & [6.105-7.212] & [5.891-7.586] & --\\
      $e$ &  & 0.8640 & 0.8556 & 0.8659 & 0.0253 & 0.8626 & [0.8410-0.8927] & [0.8239-0.9223] & [0.8087-0.9403] & --\\
      $K$ & [m/s] & 155.5 & 140.1 & 153.6 & 24.4 & 146.2 & [134.0-173.2] & [127.7-225.8] & [123.2-270.9] & --\\
      $\omega$ & [$^\circ$] & 249.20 & 252.61 & 250.16 & 4.99 & 251.29 & [245.58-254.58] & [236.70-257.20] & [231.95-259.71] & --\\
      $m$ & [$M_\oplus$] & 2101 & 2157 & 2194 & 148 & 2175 & [2059-2325] & [1954-2574] & [1856-2829] & --\\
      $m$ & [$M_\mathrm{{J}}$] & 6.610 & 6.788 & 6.905 & 0.466 & 6.843 & [6.479-7.315] & [6.148-8.098] & [5.839-8.903] & --\\
      $m$ & [$M_\odot$] & 0.006309 & 0.006479 & 0.006591 & 0.000445 & 0.006532 & [0.006184-0.006982] & [0.005868-0.007729] & [0.005574-0.008498] & --\\
      $P$ & [d] & 5633 & 6024 & 6103 & 341 & 6073 & [5769-6436] & [5506-6875] & [5274-7399] & --\\
      $T_\mathrm{C}$ & [BJD] & 2454102 & 2453718 & 2453924 & 410 & 2453860 & [2453525-2454369] & [2453243-2454828] & [2452944-2454969] & --\\
      $T_\mathrm{P}$ & [BJD] & 2455128.3 & 2455121.7 & 2455116.0 & 22.1 & 2455117.9 & [2455093.6-2455138.5] & [2455066.9-2455153.8] & [2455042.6-2455164.4] & --\\
      \hline
    \end{tabular}
  \end{center}

\end{sidewaystable*}

\begin{sidewaystable*}
\scriptsize
 \begin{center}
    \caption{Parameters probed by the MCMC used to fit the RV measurements of HD~50499.
      The maximum likelihood solution, median, mode,
      and standard-deviation of the posterior distribution
      for each parameter are shown, as well as the 68.27\%,
      95.45\%, and 99.73\% confidence intervals.
      The prior for each parameter can be of type: $\mathcal{U}$:~uniform,
      $\mathcal{N}$:~normal, or $\mathcal{TN}$:~truncated normal.
      Reference epoch: 2455500.0~BJD.}
    \label{tab:HD50499_mcmc-allparams}
     \resizebox{0.75\textwidth}{!}{
    \begin{tabular}{cccccccccccc}
      \hline
      Parameter & Units & Max(Likelihood) & Mode & Mean & Std & Median & 68.27\% & 95.45\% & 99.73\% & Prior\\
      \hline
      $\log$(Likelihood) &  & -668.22 & -677.90 & -678.84 & 3.49 & -678.50 & [-682.24--675.41] & [-686.79--672.90] & [-692.68--671.11] & --\\
        \hline
        \multicolumn{11}{c}{\textbf{Star}}\\
        \hline
      $M_\mathrm{{S}}$ & [$M_\odot$] & 1.3739 & 1.3102 & 1.3098 & 0.0699 & 1.3102 & [1.2402-1.3795] & [1.1687-1.4483] & [1.0980-1.5220] & $\mathcal{{U}}$\\
      $\Pi_\mathrm{{S}}$ & [mas] & 99.91 & 100.03 & 99.99 & 1.00 & 100.00 & [98.98-100.99] & [97.99-101.98] & [96.91-103.08] & $\mathcal{{U}}$\\
      \hline
      \multicolumn{11}{c}{\textbf{Offset}}\\
      \hline
      $\gamma_{\mathrm{HIRES(Pub-2017)}}$ & [m/s] & -36798.27 & -36798.15 & -36798.23 & 1.45 & -36798.21 & [-36799.68--36796.79] & [-36801.16--36795.38] & [-36802.59--36794.00] & $\mathcal{{U}}$\\
      $\gamma_{\mathrm{COR98(DRS-3.3)}}$ & [m/s] & 3.63 & 3.77 & 3.71 & 1.95 & 3.72 & [1.76-5.68] & [-0.21-7.57] & [-2.15-9.55] & $\mathcal{N}(0,4)$\\
      $\gamma_{\mathrm{COR07(DRS-3.4)}}$ & [m/s] & 36813.30 & 36814.53 & 36817.23 & 6.52 & 36815.30 & [36812.73-36821.25] & [36810.66-36838.24] & [36808.80-36850.47] & $\mathcal{{U}}$\\
      $\gamma_{\mathrm{HARPS03(DRS-3.5)}}$ & [m/s] & 32.74 & 31.73 & 31.76 & 2.32 & 31.75 & [29.47-34.04] & [27.06-36.43] & [24.66-39.01] & $\mathcal{N}(32,4)$\\
      $\gamma_{\mathrm{COR14(DRS-3.8)}}$ & [m/s] & 17.56 & 16.71 & 16.80 & 2.60 & 16.78 & [14.25-19.38] & [11.59-22.11] & [8.46-24.74] & $\mathcal{N}(12,4)$\\
      \hline
      \multicolumn{11}{c}{\textbf{Noise}}\\
      \hline
      $\sigma_{\mathrm{HIRES(Pub-2017)}}$ & [m/s] & 3.956 & 4.178 & 4.247 & 0.410 & 4.220 & [3.838-4.658] & [3.503-5.139] & [3.212-5.777] & $\mathcal{{U}}$\\
      $\sigma_{\mathrm{COR98(DRS-3.3)}}$ & [m/s] & 7.34 & 7.71 & 7.95 & 1.16 & 7.85 & [6.81-9.09] & [5.93-10.53] & [5.12-12.29] & $\mathcal{{U}}$\\
      $\sigma_{\mathrm{COR07(DRS-3.4)}}$ & [m/s] & 7.42 & 7.52 & 7.84 & 1.22 & 7.73 & [6.65-9.03] & [5.73-10.58] & [4.94-12.62] & $\mathcal{{U}}$\\
      $\sigma_{\mathrm{HARPS03(DRS-3.5)}}$ & [m/s] & 1.81 & 2.24 & 3.20 & 2.11 & 2.68 & [1.71-4.49] & [1.12-8.72] & [0.67-18.95] & $\mathcal{{U}}$\\
      $\sigma_{\mathrm{COR14(DRS-3.8)}}$ & [m/s] & 4.79 & 5.04 & 5.19 & 1.02 & 5.12 & [4.19-6.19] & [3.36-7.42] & [2.56-8.91] & $\mathcal{{U}}$\\
          \hline
          \multicolumn{11}{c}{\textbf{HD~50499b}}\\
          \hline
      $\log P$ & [d] & 3.39372 & 3.39533 & 3.39536 & 0.00333 & 3.39535 & [3.39207-3.39867] & [3.38870-3.40205] & [3.38489-3.40537] & $\mathcal{{U}}$\\
      $\log K$ & [m/s] & 1.2733 & 1.2783 & 1.2766 & 0.0195 & 1.2772 & [1.2573-1.2959] & [1.2361-1.3150] & [1.2148-1.3338] & $\mathcal{{U}}$\\
      $\sqrt{e}\cos\omega$ &  & -0.1092 & -0.1029 & -0.1038 & 0.0773 & -0.1033 & [-0.1813--0.0262] & [-0.2583-0.0484] & [-0.3288-0.1272] & $\mathcal{{U}}$\\
      $\sqrt{e}\sin\omega$ &  & -0.5199 & -0.5129 & -0.5060 & 0.0458 & -0.5082 & [-0.5507--0.4615] & [-0.5915--0.4069] & [-0.6298--0.3356] & $\mathcal{{U}}$\\
      $T_\mathrm{Vmin}$ & [BJD] & 2455823.0 & 2455820.7 & 2455825.2 & 33.8 & 2455823.4 & [2455791.7-2455858.5] & [2455762.5-2455898.4] & [2455733.0-2455940.5] & $\mathcal{{U}}$\\          \hline
      $a_\mathrm{S}$ & [AU] & 0.004096 & 0.004160 & 0.004153 & 0.000184 & 0.004154 & [0.003971-0.004336] & [0.003784-0.004523] & [0.003595-0.004701] & --\\
      $a$ & [AU] & 3.9830 & 3.9293 & 3.9288 & 0.0730 & 3.9298 & [3.8567-4.0018] & [3.7783-4.0719] & [3.6983-4.1391] & --\\
      $e$ &  & 0.2822 & 0.2725 & 0.2749 & 0.0382 & 0.2739 & [0.2376-0.3124] & [0.2004-0.3549] & [0.1642-0.4003] & --\\
      $K$ & [m/s] & 18.763 & 18.944 & 18.927 & 0.850 & 18.930 & [18.083-19.767] & [17.223-20.653] & [16.399-21.568] & --\\
      $\omega$ & [$^\circ$] & 258.14 & 259.32 & 258.18 & 9.08 & 258.55 & [249.13-267.21] & [238.96-275.11] & [227.46-283.12] & --\\
      $m$ & [$M_\oplus$] & 470.6 & 460.6 & 460.8 & 25.6 & 460.7 & [435.4-486.4] & [410.1-512.8] & [385.2-540.9] & --\\
      $m$ & [$M_\mathrm{{J}}$] & 1.4809 & 1.4494 & 1.4501 & 0.0807 & 1.4497 & [1.3699-1.5304] & [1.2905-1.6136] & [1.2121-1.7021] & --\\
      $m$ & [$M_\odot$] & 0.0014135 & 0.0013834 & 0.0013841 & 0.0000770 & 0.0013837 & [0.0013076-0.0014608] & [0.0012318-0.0015402] & [0.0011569-0.0016247] & --\\
      $P$ & [d] & 2475.8 & 2484.6 & 2485.3 & 19.1 & 2485.1 & [2466.5-2504.2] & [2447.4-2523.8] & [2426.0-2543.1] & --\\
      $T_\mathrm{C}$ & [BJD] & 2455055.5 & 2455060.8 & 2455055.6 & 50.9 & 2455057.1 & [2455004.9-2455106.4] & [2454950.1-2455153.9] & [2454889.1-2455194.6] & --\\
      $T_\mathrm{P}$ & [BJD] & 2457073.4 & 2456172.9 & 2456164.6 & 58.8 & 2456167.7 & [2456105.4-2456223.3] & [2456040.3-2456273.6] & [2455966.5-2456322.7] & --\\
          \hline
          \multicolumn{11}{c}{\textbf{HD~50499c}}\\
          \hline
      $\log P$ & [d] & 3.9217 & 3.9390 & 3.9819 & 0.0949 & 3.9509 & [3.9140-4.0506] & [3.8848-4.2905] & [3.8590-4.4037] & $\mathcal{{U}}$\\
      $\log K$ & [m/s] & 1.3722 & 1.3816 & 1.4081 & 0.0555 & 1.3917 & [1.3670-1.4475] & [1.3473-1.5814] & [1.3297-1.6610] & $\mathcal{{U}}$\\
      $\sqrt{e}\cos\omega$ &  & 0.086 & -0.058 & -0.103 & 0.220 & -0.089 & [-0.344-0.132] & [-0.549-0.279] & [-0.642-0.401] & $\mathcal{{U}}$\\
      $\sqrt{e}\sin\omega$ &  & 0.197 & 0.099 & 0.080 & 0.117 & 0.085 & [-0.046-0.204] & [-0.159-0.286] & [-0.256-0.349] & $\mathcal{{U}}$\\
      $T_\mathrm{Vmin}$ & [BJD] & 2454510.8 & 2454559.9 & 2454549.0 & 73.8 & 2454551.4 & [2454475.4-2454621.0] & [2454394.8-2454692.6] & [2454316.7-2454770.4] & $\mathcal{{U}}$\\          \hline
      $a_\mathrm{S}$ & [AU] & 0.0181 & 0.0191 & 0.0240 & 0.0115 & 0.0201 & [0.0177-0.0286] & [0.0161-0.0634] & [0.0148-0.0970] & --\\
      $a$ & [AU] & 8.96 & 9.02 & 9.78 & 1.68 & 9.24 & [8.69-10.75] & [8.27-15.51] & [7.88-18.57] & --\\
      $e$ &  & 0.0460 & 0.0000 & 0.0792 & 0.0761 & 0.0571 & [0.0171-0.1365] & [0.0026-0.3079] & [0.0002-0.4259] & --\\
      $K$ & [m/s] & 23.56 & 24.23 & 25.82 & 3.78 & 24.64 & [23.28-28.02] & [22.25-38.14] & [21.37-45.82] & --\\
      $\omega$ & [$^\circ$] & 66 & 180 & 50 & 113 & 81 & [-115-161] & [-177-177] & [-180-180] & --\\
      $m$ & [$M_\oplus$] & 923 & 932 & 1032 & 239 & 957 & [874-1164] & [814-1801] & [759-2346] & --\\
      $m$ & [$M_\mathrm{{J}}$] & 2.904 & 2.932 & 3.246 & 0.751 & 3.011 & [2.752-3.664] & [2.560-5.666] & [2.387-7.382] & --\\
      $m$ & [$M_\odot$] & 0.002772 & 0.002799 & 0.003098 & 0.000717 & 0.002874 & [0.002626-0.003497] & [0.002444-0.005408] & [0.002278-0.007046] & --\\
      $P$ & [d] & 8350 & 8611 & 9861 & 2745 & 8932 & [8204-11235] & [7671-19522] & [7228-25332] & --\\
      $T_\mathrm{C}$ & [BJD] & 2456009 & 2461130 & 2462211 & 2525 & 2461357 & [2460705-2463463] & [2460175-2471070] & [2459652-2476577] & --\\
      $T_\mathrm{P}$ & [BJD] & 2455509 & 2461832 & 2462479 & 4026 & 2461969 & [2458947-2465593] & [2456054-2474140] & [2455527-2479915] & --\\
      \hline
    \end{tabular}}
  \end{center}

\end{sidewaystable*}

\begin{sidewaystable*}
%\scriptsize
  \begin{center}
    \caption{Parameters probed by the MCMC used to fit the RV measurements of HD~92788.
      The maximum likelihood solution, median, mode,
      and standard-deviation of the posterior distribution
      for each parameter are shown, as well as the 68.27\%,
      95.45\%, and 99.73\% confidence intervals.
      The prior for each parameter can be of type: $\mathcal{U}$:~uniform,
      $\mathcal{N}$:~normal, or $\mathcal{TN}$:~truncated normal.
      Reference epoch: 2455500.0~BJD.}
    \label{tab:HD92788_mcmc-allparams}
    \resizebox{0.75\textwidth}{!}{
    \begin{tabular}{cccccccccccc}
      \hline
      Parameter & Units & Max(Likelihood) & Mode & Mean & Std & Median & 68.27\% & 95.45\% & 99.73\% & Prior\\
      \hline
      $\log$(Likelihood) &  & -625.86 & -636.46 & -637.60 & 3.58 & -637.24 & [-641.13--634.06] & [-645.81--631.54] & [-650.84--629.55] & --\\
        \hline
        \multicolumn{11}{c}{\textbf{Star}}\\
        \hline
      $M_\mathrm{{S}}$ & [$M_\odot$] & 1.1684 & 1.1483 & 1.1500 & 0.0702 & 1.1500 & [1.0805-1.2202] & [1.0089-1.2907] & [0.9272-1.3575] & $\mathcal{{U}}$\\
      $\Pi_\mathrm{{S}}$ & [mas] & 28.838 & 28.835 & 28.829 & 0.288 & 28.828 & [28.541-29.120] & [28.254-29.406] & [27.958-29.662] & $\mathcal{{U}}$\\
      \hline
      \multicolumn{11}{c}{\textbf{Offset}}\\
      \hline
      $\gamma_{\mathrm{COR14(DRS-3.8)}}$ & [m/s] & 9.64 & 10.96 & 10.84 & 3.53 & 10.83 & [7.32-14.39] & [3.89-17.94] & [0.33-21.77] & $\mathcal{N}(12,4)$\\
      $\gamma_{\mathrm{COR07(DRS-3.4)}}$ & [m/s] & -4430.37 & -4430.03 & -4430.76 & 2.76 & -4430.42 & [-4433.39--4428.15] & [-4437.43--4426.17] & [-4440.86--4423.54] & $\mathcal{{U}}$\\
      $\gamma_{\mathrm{HAMILTON(Pub-2006)}}$ & [m/s] & 4448.46 & 4448.30 & 4448.04 & 2.68 & 4448.10 & [4445.39-4450.72] & [4442.43-4453.25] & [4439.52-4455.84] & $\mathcal{{U}}$\\
      $\gamma_{\mathrm{COR98(DRS-3.3)}}$ & [m/s] & -1.95 & -2.24 & -2.24 & 2.29 & -2.21 & [-4.51-0.03] & [-6.94-2.25] & [-9.77-4.46] & $\mathcal{N}(0,4)$\\
      $\gamma_{\mathrm{HARPS03(DRS-3.5)}}$ & [m/s] & 38.84 & 38.53 & 38.23 & 1.84 & 38.33 & [36.45-40.02] & [34.19-41.68] & [31.45-43.36] & $\mathcal{N}(32,4)$\\
      $\gamma_{\mathrm{HIRES(Pub-2017)}}$ & [m/s] & 4470.60 & 4470.24 & 4470.10 & 1.93 & 4470.16 & [4468.25-4472.01] & [4465.96-4473.73] & [4462.84-4475.55] & $\mathcal{{U}}$\\
      \hline
      \multicolumn{11}{c}{\textbf{Noise}}\\
      \hline
      $\sigma_{\mathrm{COR07(DRS-3.4)}}$ & [m/s] & 2.42 & 3.96 & 4.54 & 2.72 & 4.24 & [1.90-6.99] & [0.33-10.98] & [0.02-17.75] & $\mathcal{{U}}$\\
      $\sigma_{\mathrm{HIRES(Pub-2017)}}$ & [m/s] & 3.521 & 3.752 & 3.831 & 0.531 & 3.788 & [3.311-4.355] & [2.911-5.025] & [2.576-5.898] & $\mathcal{{U}}$\\
      $\sigma_{\mathrm{COR14(DRS-3.8)}}$ & [m/s] & 2.91 & 4.08 & 4.60 & 2.14 & 4.36 & [2.65-6.52] & [0.96-9.71] & [0.07-14.94] & $\mathcal{{U}}$\\
      $\sigma_{\mathrm{HAMILTON(Pub-2006)}}$ & [m/s] & 4.85 & 5.16 & 5.08 & 2.22 & 5.09 & [2.82-7.25] & [0.62-9.64] & [0.03-12.51] & $\mathcal{{U}}$\\
      $\sigma_{\mathrm{COR98(DRS-3.3)}}$ & [m/s] & 9.36 & 9.58 & 9.83 & 1.35 & 9.75 & [8.49-11.17] & [7.40-12.78] & [6.42-14.65] & $\mathcal{{U}}$\\
      $\sigma_{\mathrm{HARPS03(DRS-3.5)}}$ & [m/s] & 1.431 & 1.491 & 1.535 & 0.194 & 1.523 & [1.344-1.727] & [1.186-1.960] & [1.045-2.215] & $\mathcal{{U}}$\\
          \hline
          \multicolumn{11}{c}{\textbf{HD~92788b}}\\
          \hline
    $\log P$ & [d] & 2.5128215 & 2.5128306 & 2.5128344 & 0.0000622 & 2.5128330 & [2.5127722-2.5128955] & [2.5127131-2.5129638] & [2.5126589-2.5130376] & $\mathcal{{U}}$\\
    $\log K$ & [m/s] & 2.03396 & 2.03454 & 2.03446 & 0.00351 & 2.03447 & [2.03100-2.03795] & [2.02741-2.04146] & [2.02355-2.04503] & $\mathcal{{U}}$\\ 
    $\sqrt{e}\cos\omega$ &  & 0.0809 & 0.0797 & 0.0800 & 0.0113 & 0.0800 & [0.0687-0.0912] & [0.0573-0.1026] & [0.0452-0.1144] & $\mathcal{{U}}$\\ 
      $\sqrt{e}\sin\omega$ &  & -0.58785 & -0.58698 & -0.58697 & 0.00398 & -0.58699 & [-0.59092--0.58302] & [-0.59485--0.57895] & [-0.59886--0.57446] & $\mathcal{{U}}$\\
      $\lambda_0$ & [$^\circ$] & 115.191 & 115.166 & 115.125 & 0.474 & 115.140 & [114.656-115.591] & [114.128-116.041] & [113.540-116.525] & $\mathcal{{U}}$\\        \hline
      $a_\mathrm{S}$ & [AU] & 0.0030300 & 0.0030350 & 0.0030349 & 0.0000271 & 0.0030348 & [0.0030080-0.0030619] & [0.0029814-0.0030896] & [0.0029521-0.0031194] & --\\
      $a$ & [AU] & 0.9768 & 0.9720 & 0.9712 & 0.0198 & 0.9716 & [0.9517-0.9910] & [0.9302-1.0097] & [0.9044-1.0267] & --\\
      $e$ &  & 0.35211 & 0.35120 & 0.35108 & 0.00451 & 0.35108 & [0.34660-0.35558] & [0.34200-0.36013] & [0.33750-0.36478] & --\\
      $K$ & [m/s] & 108.133 & 108.243 & 108.262 & 0.874 & 108.261 & [107.400-109.131] & [106.516-110.016] & [105.571-110.926] & --\\
      $\omega$ & [$^\circ$] & -82.17 & -82.27 & -82.24 & 1.10 & -82.24 & [-83.33--81.16] & [-84.44--80.04] & [-85.61--78.90] & --\\
      $m$ & [$M_\oplus$] & 1208.0 & 1195.9 & 1196.7 & 50.0 & 1196.9 & [1147.2-1247.0] & [1095.5-1295.0] & [1037.4-1343.4] & --\\
      $m$ & [$M_\mathrm{{J}}$] & 3.801 & 3.763 & 3.766 & 0.157 & 3.766 & [3.610-3.924] & [3.447-4.075] & [3.264-4.227] & --\\
      $m$ & [$M_\odot$] & 0.003628 & 0.003592 & 0.003594 & 0.000150 & 0.003595 & [0.003446-0.003745] & [0.003290-0.003890] & [0.003116-0.004035] & --\\
      $P$ & [d] & 325.7028 & 325.7097 & 325.7125 & 0.0467 & 325.7114 & [325.6659-325.7583] & [325.6215-325.8095] & [325.5809-325.8649] & --\\
      $T_\mathrm{C}$ & [BJD] & 2455470.50 & 2455470.61 & 2455470.65 & 1.29 & 2455470.64 & [2455469.38-2455471.93] & [2455468.11-2455473.28] & [2455466.78-2455474.73] & --\\
      $T_\mathrm{P}$ & [BJD] & 2455543.031 & 2455647.141 & 2455647.142 & 0.736 & 2455647.139 & [2455646.415-2455647.868] & [2455645.674-2455648.640] & [2455644.949-2455649.443] & --\\
          \hline
          \multicolumn{11}{c}{\textbf{HD~92788c}}\\
          \hline
    $\log P$ & [d] & 4.077 & 4.065 & 4.122 & 0.107 & 4.094 & [4.030-4.222] & [3.984-4.410] & [3.945-4.597] & $\mathcal{{U}}$\\
    $\log K$ & [m/s] & 1.5369 & 1.5238 & 1.5239 & 0.0278 & 1.5239 & [1.4962-1.5517] & [1.4684-1.5805] & [1.4393-1.6035] & $\mathcal{{U}}$\\
    $\sqrt{e}\cos\omega$ &  & 0.6302 & 0.6139 & 0.6196 & 0.0554 & 0.6185 & [0.5676-0.6730] & [0.5065-0.7354] & [0.4308-0.7870] & $\mathcal{{U}}$\\
      $\sqrt{e}\sin\omega$ &  & -0.2734 & -0.3134 & -0.3152 & 0.0947 & -0.3135 & [-0.4112--0.2213] & [-0.5121--0.1313] & [-0.5867--0.0255] & $\mathcal{{U}}$\\
      $T_\mathrm{Vmax}$ & [BJD] & 2457168.9 & 2457142.6 & 2457122.4 & 99.5 & 2457129.7 & [2457025.2-2457220.2] & [2456899.1-2457298.5] & [2456754.7-2457390.4] & $\mathcal{{U}}$\\          \hline
      $a_\mathrm{S}$ & [AU] & 0.03328 & 0.03195 & 0.03578 & 0.00782 & 0.03359 & [0.02964-0.04200] & [0.02686-0.05768] & [0.02452-0.08454] & --\\
      $a$ & [AU] & 10.77 & 10.50 & 11.65 & 2.18 & 11.00 & [9.95-13.40] & [9.24-17.84] & [8.65-23.92] & --\\
      $e$ &  & 0.4719 & 0.4551 & 0.4952 & 0.0767 & 0.4793 & [0.4241-0.5739] & [0.3805-0.6820] & [0.3427-0.7579] & --\\
      $K$ & [m/s] & 34.43 & 33.29 & 33.48 & 2.14 & 33.41 & [31.35-35.62] & [29.40-38.06] & [27.50-40.13] & --\\
      $\omega$ & [$^\circ$] & -23.45 & -25.71 & -26.79 & 7.89 & -26.49 & [-34.63--19.08] & [-43.21--11.50] & [-52.07--2.23] & --\\
      $m$ & [$M_\oplus$] & 1203.2 & 1164.8 & 1172.1 & 87.7 & 1169.0 & [1084.5-1260.1] & [1004.8-1356.5] & [929.2-1455.6] & --\\
      $m$ & [$M_\mathrm{{J}}$] & 3.786 & 3.665 & 3.688 & 0.276 & 3.678 & [3.413-3.965] & [3.162-4.268] & [2.924-4.580] & --\\
      $m$ & [$M_\odot$] & 0.003614 & 0.003498 & 0.003520 & 0.000264 & 0.003511 & [0.003257-0.003785] & [0.003018-0.004074] & [0.002791-0.004372] & --\\
      $P$ & [d] & 11927 & 11610 & 13703 & 4106 & 12405 & [10706-16666] & [9639-25718] & [8804-39575] & --\\
      $T_\mathrm{C}$ & [BJD] & 2461454 & 2458838 & 2458997 & 386 & 2458921 & [2458666-2459320] & [2458452-2460040] & [2458240-2460932] & --\\
      $T_\mathrm{P}$ & [BJD] & 2456919 & 2456858 & 2456825 & 167 & 2456834 & [2456656-2456991] & [2456473-2457133] & [2456281-2457257] & --\\
      \hline
    \end{tabular}}
  \end{center}
\end{sidewaystable*}

\end{appendix}

\end{document}